\documentclass[journal,10pt,twocolumn]{IEEEtran}
\usepackage{mathrsfs}
\usepackage{graphicx}
\usepackage{subfigure}
\usepackage{multirow}
\usepackage{amsfonts}
\usepackage{amssymb}
\usepackage{amsmath}
\usepackage{url}
\usepackage{lineno}
\usepackage{latexsym}
\usepackage{amsxtra}
\usepackage{array}
\usepackage{color}
\usepackage{subfigure}
\usepackage{algorithmic,algorithm}
\usepackage{enumerate}
\usepackage{cite}
\usepackage{bbm}
\newtheorem{theorem}{Theorem}
\newtheorem{lemma}{Lemma}

\newtheorem{definition}{Definition}

\begin{document}
	
	\title{ Two-Stage Channel Estimation Approach for Cell-Free IoT With Massive Random Access}
	
	\author{ Xinhua Wang,  \textit{Member, IEEE},  Alexei Ashikhmin, \textit{Fellow, IEEE},  
		
		Zhicheng Dong, \textit{Member, IEEE}, and Chao Zhai, \textit{Member, IEEE}  
		
		\thanks{Copyright (c) 2015 IEEE. Personal use of this material is permitted. However, permission to use this material for any other purposes must be obtained from the IEEE by sending a request to pubs-permissions@ieee.org.
		
X. Wang is with the College of Electrical Engineering, Qingdao University, Qingdao, 266071 China (e-mail: xhwang@qdu.edu.cn).
         
A. Ashikhmin is with the Nokia Bell Labs, Murray Hill, NJ 07974 USA (e-mail: alexei.ashikhmin@nokia-bell-labs.com).   
 
 Z. Dong is with the School of Engineering, Tibet University, Lhasa, 850000 China (e-mail: dongzc666@163.com).
 
 C. Zhai is with the School of Information Science and Engineering, Shandong University, Qingdao, 266237 China (e-mail: chaozhai@sdu.edu.cn).

 }
	}
	
	\maketitle
	
	\begin{abstract}
We investigate the activity detection and channel estimation issues for cell-free Internet of Things (IoT) networks with massive random access. In each time slot, only partial devices are active and communicate with neighboring access points (APs) using non-orthogonal random pilot sequences. Different from the centralized processing in cellular networks, the activity detection and channel estimation in cell-free IoT is more challenging due to the distributed and user-centric architecture. We propose a two-stage approach to detect the random activities of devices and estimate their channel states. In the first stage, the activity of each device is jointly detected by its adjacent APs based on the vector approximate message passing (Vector AMP) algorithm. In the second stage, each AP re-estimates the channel using the linear minimum mean square error (LMMSE) method based on the detected activities to improve the channel estimation accuracy. We derive closed-form expressions for the activity detection error probability and the mean-squared channel estimation errors for a typical device. Finally, we analyze the performance of the entire cell-free IoT network in terms of coverage probability. Simulation results validate the derived closed-form expressions and show that the cell-free IoT significantly outperforms the collocated massive MIMO and small-cell schemes in terms of coverage probability. 
\end{abstract}
	
	\section{Introduction}
		\IEEEPARstart{I}{nternet} of Things (IoT) is a promising technology to enable massively connected intelligent objects to make decisions cooperatively. IoT has been wildly exploited in many fields, such as smart healthcare and intelligent transportation, which can bring revolutionary changes to our daily life \cite{A. Al-Fuqaha}. With the increasing development of IoT, the number of machine-type connections has grown exponentially with a rate of tens of billions a year \cite{S. Mumtaz}. Different with human-driven communications, the data traffic of machine-type communications (MTC) is often light, but requires random access to the network. Typically, in each given time slot, only a small fraction of devices are active, which shows the nature of sparsity \cite{K. Senel}.
	
	In order to support the massive connectivity, a number of coordinated access and user scheduling schemes have been proposed \cite {M. Hasan, D. Zhai}. Hasan \textit{et al.} proposed a random access mechanism similarly as the classic ALOHA, wherein each active device first randomly selects a pilot sequence from a finite set and transmits it to notify the base station (BS) about its status \cite {M. Hasan}. After receiving the response from a BS, the active device sends a data transmission request, which may be denied if it picks the same pilot sequence with another active device. With the help of Lyapunov optimization,  Zhai \textit{et al.} designed an energy-efficient user scheduling and power allocation scheme for the NOMA based IoT, which aims to minimize the network power consumption while meeting the long-term rate requirements of all the devices \cite {D. Zhai}. Unfortunately, the extra control signaling significantly increases the overhead of IoT networks, which may be even heaver than the data traffic. 
	
	Apart from the grant-based access schemes mentioned above, the grant-free IoT with random pilots has also been attracting intensive research interests, as each device can directly transmit its pilot and data to the BS without waiting for any permission \cite{L. Liu_3}. Thanks to the sparsity of active users, the compressed sensing (CS) techniques can be adopted for the activity detection. With perfect channel state information (CSI) available, a block-wise orthogonal least-squares detection was proposed in \cite{H. Zhu} to jointly detect user activity and data. Leveraging the Gaussian distribution of random pilots, the approximate message passing (AMP) algorithm can be used to jointly detect the user activity and estimate the CSI. For a single BS with single antenna, the joint activity and channel estimation problem becomes a CS single measurement vector problem, which can be efficiently solved using the AMP algorithm \cite{Z. Chen_1}. For a single BS with multiple antennas, a CS multiple measurement vector problem can be formulated as in \cite{Z. Chen_2}, which can be solved  using the similar arguments. Through asymptotic analysis, the detection error probability of the AMP-based algorithm were derived for a single-cell massive MIMO system in \cite{L. Liu_1}. Due to the severe path-loss, cell-boundary devices often suffer from very poor detection probability.

	Recently, cell-free massive MIMO has been proposed as a promising technique for the next-generation wireless systems with many distributed APs cooperatively transmitting data \cite{H. Q. Ngo_a}. In contrast to cellular networks, each user is served jointly by neighboring APs \cite{J. Zhang}. As a result, the severe path-loss effect can be alleviated in the distributed and user-centric architecture \cite{S. Buzzi}. In order to acquire the CSI, each AP can locally estimate the CSI using the linear minimum mean square error (LMMSE) technique. Considering the massive connectivity in IoT networks, the optimal LMMSE channel estimation with non-orthogonal random pilots were studied in \cite{S_Rao} and \cite{Xinhua Wang1} for single and multiple antenna cases, respectively. Using the local CSI, each AP pre-processes the received signals separately, and then delivers the generated scalars to the CPU for the combination and decoding. In all of these works, it is assumed that the active users are all correctly identified, thus APs need to estimate the channel coefficients of all the active users. This assumption is not suitable for the grant-free IoT networks with massive connectivity.

	
	In this work, we consider a user-centric and distributed cell-free network, wherein each device is served by all the APs located in a circular area with a given radius around the device. Activity detection in cell-free networks is a challenging problem, because the previously used CS techniques require centralized processing at the cellular BS, which is not suitable for the user-centric cell-free network \cite{A.-S. Bana}. Inspired by this fact, we propose a two-stage approach to detect active users and estimate their channel coefficients, and further analyze the system performance. Our main contributions are three-fold:
	\begin{itemize}
		\item We properly model the stochastic cell-free IoT network with massive random access in the $\mathbb{R}^2$ regime. We propose a two-stage approach to detect active users and estimate their channel coefficients. In the first stage, the activity of each device is jointly determined by its adjacent APs based on the vector AMP algorithm. In the second stage, the APs re-estimate the CSI using the low-complexity LMMSE method to improve the accuracy of CSI estimation for the active users.  
		\item For each device, we derive closed-form expressions for the activity detection error probability and the mean-squared channel estimation error through asymptotic analysis. Furthermore, we study the impact of the number of antennas applied at each AP, and the density of APs. 

		\item  For the entire stochastic network, we analyze the coverage probability and make comparisons with the cellular massive MIMO and small-cells. Simulation results verify the accuracy of the derived closed-form expressions and show that the user-centric approach can significantly improve the accuracy of channel estimation.
	\end{itemize}

 The rest of this paper is organized as follows. In Section II, we describe the system model and outline our
 results. In Section III, we illustrate the proposed two-stage approach. Section IV and V analyze the performance of device and network, respectively. Simulation results are provided in Section VI. Finally, Section VII concludes this paper.

\emph{\textbf{Notation}}: Throughout this paper, scalars and vectors are denoted by lowercase letters and boldface lowercase letters, respectively. $\left|\cdot\right|$ and $\left\|\cdot\right\|$ represent the absolute value and the $\ell_2$ norm, respectively. $(\cdot)^H$ and $(\cdot)^{-1}$ denote the conjugate transpose and the inverse operation, respectively.  $\mathcal{CN}\left(\textbf{m},\textbf{R}\right)$ denotes the circularly symmetric complex Gaussian (CSCG) distribution with mean vector $\textbf{m}$ and covariance matrix $\textbf{R}$. $\mathbb{E}[\cdot]$ and var$\{\cdot\}$ stand for expectation and variance operations, respectively. $\Gamma(\cdot)$, $\gamma(\cdot,\cdot)$, $\Gamma(\cdot,\cdot)$ denote the Gamma function, the lower incomplete Gamma function, and the upper incomplete
Gamma function, respectively. $x\sim {\rm Gamma}(a,b)$ represents that a random variable $x$ is gamma-distributed with shape $a$ and scale $b$.

\section{System Model and Outline of Results}
\begin{figure}
\centering \scalebox{1}{\includegraphics[width=\columnwidth]{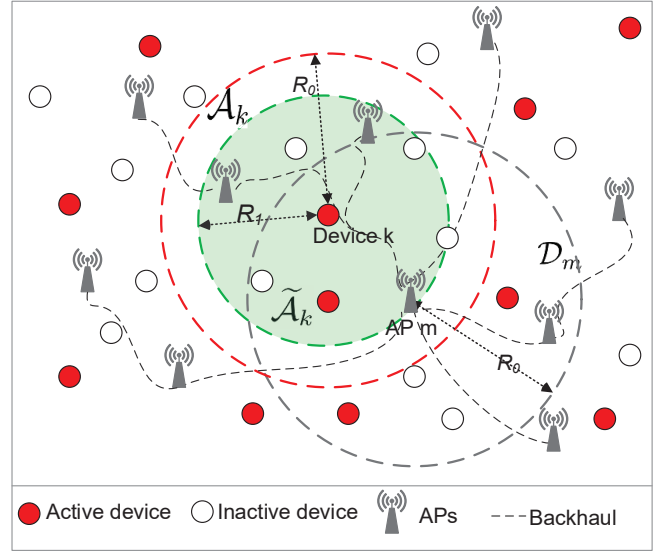}}
\centering \caption{The stochastic cell-free IoT network with massive random access. The gray circle with radius $R_0$ is the coverage area of the $m$-th AP, and $\mathcal{D}_m$ is the set of the devices within the gray circle. The red circle with radius $R_0$ is the coverage area of the  $k$-th device, and $\mathcal{A}_k$ is the set of the APs within the red circle. The activity of the $k$-th device is jointly determined by the APs denoted as $\widetilde{\mathcal{A}}_k$ within the green circle with cooperation radius $R_1<R_0$.}
\end{figure}
\subsection{System Model}
	As shown in Fig. 1, we consider a stochastic user-centric cell-free IoT network, which consists of randomly distributed APs and IoT devices. Each AP is equipped with \(N\) antennas and connects to a central processing unit (CPU), via a  back-haul network. We assume that the locations of APs follow a homogeneous Poisson point process (PPP) ${\Phi}_A$ with density \(\lambda_{\rm A}\), while the locations of devices follow another independent homogeneous PPP ${\Phi}_D$ with higher density \(\lambda_{\rm D}\rightarrow \infty\). In each time slot, each device independently transmit data to APs with probability \(\epsilon\). Let $r_{m,k}$ denote the distance between the $m$-th AP and the $k$-th device. Let ${\pmb g}_{m,k}\in \mathbb{C}^{N\times 1}$ denote the corresponding channel coefficients. Unlike a cellular network, each device is served by its adjacent APs within its {\em coverage area}. Here, the {\em coverage area} of each AP or each device is a circular area with sufficient large radius \(R_0\). The \(n\)-th element of ${\pmb g}_{m,k}$ is modeled as
\begin{align*}
    g_{(m,n),k}=\left\{\begin{array}{ll}\sqrt{\beta_{m,k}}h_{(m,n),k}, &  {\rm if}~r_{{m},k}\leq R_0, \\
0, & \mbox{otherwise},
\end{array}\right. 
\end{align*}
where \(h_{(m,n),k}\sim \mathcal{CN}(0,1) \) is the small-scale fading, and $\sqrt{\beta_{m,k}}$ denotes the path-loss which decreases monotonically with $r_{m,k}$. As shown in Fig. 1, the $k$-th device is served by a set of APs denoted by $\mathcal{A}_k$, which are located within its coverage area denoted by a red circle. Let \(\mathcal{D}_m\) be the set of devices within the coverage area of the \(m\)-th AP denoted by the gray circle.

Let \(\alpha_{k}\) be the activity indicator of the $k$-th device, i.e.,
\begin{align}\label{activity_definition}
    {\alpha}_{k}=\left\{\begin{array}{ll}1, & {\rm if} ~k\mbox{-th device is active}, \\
0, & \mbox{otherwise}.
\end{array}\right. 
\end{align}
Let \({\Pr}(\alpha_{k}=1)=\epsilon, \forall k\). We assume that the activity detection for the $k$-th device is conducted by APs located within the circular area of radius $R_1\le R_0$  (the green circle in Fig. 1). We call $R_1$ the {\em cooperative radius}. We denote the cooperative APs by the set $\widetilde{\mathcal{A}}_k=\{m:r_{m,k}\leq R_1\}$. 

In the channel estimation phase, each active device transmits the pilot sequence \(\pmb{\psi}_{k} \sim \mathcal{CN}\left( 0, \frac{1}{\tau}{\pmb I}_{\tau}\right)\) with a normalized power \(\rho\), where $\tau$ is the length of each pilot sequence.  For the \(m\)-th AP, we define \({\pmb \alpha}_m={\rm diag}(\cdots, \alpha_{k},\cdots)\), \({\pmb \beta}_m={\rm diag}(\cdots, \beta_{m,k},\cdots)\), and \(\pmb{\Psi}_m=\left[\cdots, \pmb{\psi}_{k},\cdots\right]\sim \mathbb{C}^{\tau\times |\mathcal{D}_m|}\) where $k\in \mathcal{D}_m$, and \(\pmb{\Psi}_m\) and \({\pmb \beta}_m\) are assumed to be available for the \(m\)-th AP. Let \(E=\tau\rho\), the received signal at the \(n\)-th antenna of the \(m\)-th AP is given by
	\begin{align}
	\pmb{y}_{(m,n)}&=\sqrt{E}\sum\nolimits_{k\in \mathcal{D}_m}\alpha_{k}{g}_{(m,n),k}\pmb{\psi}_{k}+{\pmb{w}}_{(m,n)}\nonumber\\
	&=\sqrt{E}\pmb{\Psi}_m{\pmb \alpha}_m\pmb{g}_{(m,n)}+{\pmb{w}}_{(m,n)},\label{receivd_pilot_EAN}
	\end{align}
	where \({\pmb{w}}_{(m,n)}\sim \mathcal{CN}(0,\pmb{I}_{\tau})\) is the normalized noise, and
	\begin{align}
	    \pmb{g}_{(m,n)}=\left[\cdots,g_{(m,n),k},\cdots\right]^T ~\in \mathbb{C}^{|\mathcal{D}_m| \times 1}, ~k\in \mathcal{D}_m  \nonumber
	\end{align}
	denotes the channel coefficients vector between the \(n\)-th antenna of the \(m\)-th AP and the devices in \(\mathcal{D}_m\). The channel coefficients between the \(m\)-th AP and the $k$-th device are given by
	\begin{align}\label{g_mk}
	    \pmb{g}_{m,k}=\left[g_{(m,1),k},\cdots,g_{(m,N),k}\right]^T ~\in \mathbb{C}^{N \times 1}, ~k\in \mathcal{D}_m, 
	\end{align}
with $\pmb{g}_{m,k}\sim \mathcal{CN}(0,\beta_{m,k}{\pmb I}_{N}),~k\in \mathcal{D}_m$. We combine the channel coefficients between all the $N$ antennas of the $m$-th AP and all the devices in \(\mathcal{D}_m\) into the matrix  
	\[{\pmb{G}}_m=[{{\pmb{g}}}_{(m,1)}, \cdots, \pmb{g}_{(m,N)}] =[\cdots, {{\pmb{g}}}_{m,k}, \cdots]^T,~k\in \mathcal{D}_m.\] 
According to (\ref{receivd_pilot_EAN}), the received pilots at the \(m\)-th AP can be expressed as 
 \begin{align}
\pmb{Y}_m&=\sqrt{E}\pmb{\Psi}_m\pmb{\alpha}_m\pmb{G}_m+{\pmb{W}}_m=\sqrt{E}\pmb{\Psi}_m\pmb{X}_m+{\pmb{W}}_m, \label{received_Pilot_EAP}
\end{align}	
where \[\pmb{Y}_m=[\pmb{y}_{(m,1)},\cdots,\pmb{y}_{(m,N)}],\] \[\pmb{W}_m=[\pmb{w}_{(m,1)},\cdots,\pmb{w}_{(m,N)}],\] and
\[\pmb{X}_m=\pmb{\alpha}_m\pmb{G}_m=[\cdots,\pmb{x}_{m,k},\cdots]^T,~k\in \mathcal{D}_m,\] 
with \begin{align}
{\pmb x}_{m,k}=\alpha_{k}{\pmb g}_{m,k}.
\end{align}
According to (\ref{activity_definition}), the distribution of ${\pmb x}_{m,k}$ is 
\begin{equation}\label{pdf_channel} 
p_{{\pmb x}_{m,k}}=(1-\epsilon)\delta_{\bf 0}+\epsilon p_{{\pmb g}_{m,k}},
\end{equation}
where $\delta_{\bf 0}$ denotes the point mass measure at zero as the $k$-th device is inactive, and $p_{{\pmb g}_{m,k}}$ is the probability density function (PDF) of $\pmb{g}_{m,k}$ defined in (\ref{g_mk}).

For the ease of understanding, some notations of this paper are summarized in Table I. To facilitate analysis, we have the following assumptions made throughout this paper:
\begin{itemize}
	\item We consider the cell-free IoT with massive connectivity, i.e., \(\lambda_D\rightarrow\infty\), which implies the number of active devices within the coverage of $m$-th AP $|\Omega_m|/|\mathcal{D}_m|\rightarrow \epsilon $, ~$\forall m$.
	\item Similar to \cite{L. Liu_1}, we consider the asymptotic regime with \(T,\tau,|\mathcal{D}_m|\rightarrow \infty\), while keeping the total transmit energy of pilot \(E=\tau\rho\), and the ratios \(\tau/T\) and \(\tau/|\mathcal{D}_m|\) fixed. 
\end{itemize}

\begin{table}[t!]
	\renewcommand{\arraystretch}{1.3}
	\caption{}
	\label{table_parameters} \vskip-2mm
	\centering
	\begin{tabular}{p{1cm}p{7cm}}
		\hline
		\bfseries Notation & \bfseries Meaning \\
		\hline $\Phi_A$ & Poisson
		point process (PPP) of APs \\
		\hline $\lambda_A$ & Density of $\Phi_A$\\
		\hline $\Phi_D$ & PPP of devices \\
		\hline $\lambda_D$ & Density of $\Phi_D$\\
		\hline $r_{m,k}$ &  Distance between the $m$-th AP and the $k$-th device\\
		\hline $\alpha_k$ &  Activity indicator of the $k$-th device\\
		\hline $\widehat{\alpha}_k$ & estimate of $\alpha_k$\\
		\hline ${\pmb \psi}_k$ &  Pilot sequence of the $k$-th device\\
		\hline $\tau$ & Length of pilot \\
		\hline $E$ & Total transmit energy of pilot, i.e., $E=\tau\rho$\\
		\hline $\rho$ & Normalized pilot transmit power \\
		\hline $R_0$ & Coverage radius \\
		\hline $R_1$ & Cooperation radius\\
		\hline $\mathcal{A}_k$ & Set of APs within the coverage radius of the $k$-th device, i.e., $\mathcal{A}_k=\{m:r_{m,k}\leq R_0\}$.\\
		\hline $\widetilde{\mathcal{A}}_k$ & Set of APs to jointly determine $\alpha_k$, i.e., $\widetilde{\mathcal{A}}_k=\{m:r_{m,k}\leq R_1\}$ \\
		\hline $\mathcal{D}_m$ & Set of devices within the coverage radius of the $m$-th AP, i.e.,  ${\mathcal{D}}_m=\{k:d_{m,k}\leq R_0\}$\\
		\hline $\Omega_m$ & Active devices within the coverage radius of the $m$-th AP, i.e., $\Omega_m=\{k:\alpha_{k}=1, k\in {\mathcal{D}}_m\}$\\
		\hline
	\end{tabular}
\end{table}


For the $m$-th AP, the path-loss coefficients $\beta_{m,k}, k\in \mathcal{D}_m$ follow independent identical distribution for different geometry realizations, i.e., $\beta_{m,k}\sim p_{\beta}, \forall k\in \mathcal{D}_m$. For a snap shot, each geometry realization of \(\{\beta_{m,k}:k\in \mathcal{D}_m\}\) is equivalent to $|\mathcal{D}_m|$ multiplied by the realization of any $\beta_{m,k}$ with $k\in \mathcal{D}_m$. Thus, the PDF of each geometry realization of \(\{\beta_{m,k}:k\in \mathcal{D}_m\}\) converges to $p_{\beta}$ as $\lambda_D\rightarrow \infty$.

\subsection{Outline of Results}


For the distributed and user-centric architecture, it is nontrivial to detect the user activity and estimate the CSI for cell-free IoT networks due to the following reasons: (1) The activity $\alpha_k$ is jointly determined by its adjacent APs, which means the traditional centralized detection techniques \cite{L. Liu_1} for cellular networks are unavailable. (2) Considering the overhead over back-haul, the distributed APs cannot fully cooperate to detect user activities. (3) Compared with \cite{Xinhua Wang1}, the performance analysis for LMMSE channel estimation is more challenging due to the possible errors of activity detection. 

In Section III, we will propose a two-stage channel estimation approach based on the AMP algorithm and the LMMSE method. 
\begin{itemize}
    \item Stage I: According to the signals \(\pmb{Y}_m\) defined in (\ref{received_Pilot_EAP}), each AP first estimates \(\pmb{X}_m\) using the AMP algorithm. Then, the activity $\widehat{\alpha}_{k}$ of the $k$-th device is jointly determined by fusing the estimates \(\{\widehat{\pmb{x}}_{m,k}:m\in \widetilde{\mathcal{A}}_k\}\) from the adjacent APs $\widetilde{\mathcal{A}}_k$.
    \item Stage II: The channel vectors \(\{\widehat{\pmb g}_{m,k}:k\in \mathcal{D}_m\}\) are re-estimated by the \(m\)-th AP using the low-complexity LMMSE method based on \(\widehat{\pmb \alpha}_m={\rm diag}(\cdots, \alpha_{k},\cdots), k\in \mathcal{D}_m\) and \(\pmb{Y}_m\).
\end{itemize}

In Section IV, the activity detection error probability and the mean-squared error of channel estimate are derived in closed-form using the asymptotic analysis. In addition, we investigate how the number of antennas and the density of APs affect the system performance. 

In Section V, we define the coverage probability as a metric to evaluate the performance for the entire stochastic cell-free IoT network. Then, we compare our user-centric based channel estimation approach with the collocated massive MIMO and small-cell in terms of coverage probability .

\section{Two-stage Channel Estimation Approach}


 
\subsection{Pre-processing at Each AP for Stage I}
First, the $m$-th AP independently estimates \(\pmb{X}_m\) based on \(\pmb{Y}_m\) using the vector AMP algorithm. Similarly to \cite{L. Liu_1}, the vector AMP algorithm is updated as follows with initialization $\pmb{X}_{m,{(0)}}=\pmb{0}$ and $\pmb{R}_{m,{(0)}}=\pmb{Y}_m$,
\begin{align}
\pmb{x}_{{m,k},{(t+1)}} & =\eta\left(\widehat{\pmb{x}}_{{m,k},{(t)}}\right), k\in \mathcal{D}_m, \label{eqn:AMP 1} \\
\pmb{R}_{m,{(t+1)}} & =\pmb{Y}_m-\pmb{\Psi}_m\pmb{X}_{m,{(t+1)}}\nonumber\\
& +\frac{|\mathcal{D}_m|}{\tau}\pmb{R}_{m,{(t)}}\sum\nolimits_{k\in \mathcal{D}_m}\frac{\eta'(\widehat{\pmb{x}}_{{m,k},{(t)})}}{|\mathcal{D}_m|}, \label{eqn:AMP 2}
\end{align}
where $t=0,1,\cdots$ is the index of iteration, and  
$$\widehat{\pmb{x}}_{{m,k},{(t)}}=\pmb{R}_{m,{(t)}}^H\pmb{\psi}_{m,k}+\pmb{x}_{{m,k},{(t)}}.$$
$\pmb{X}_{m,{(t)}}=[\cdots,\pmb{x}_{{m,k},{(t)}},\cdots]^T, ~k\in\mathcal{D}_m$ is the estimate of $\pmb{X}_m$ at the $t$-th iteration, and $\pmb{R}_{m,{(t)}}=[\cdots,\pmb{r}_{{m,k},{(t)}},\cdots]^T, ~k\in\mathcal{D}_m$ represents the corresponding residual. \(\eta(\cdot)\) and \(\eta'(\cdot)\) represent an appropriately designed denoiser and its first-order derivative, respectively. Using the state evolution \cite{L. Liu_1}, \(\widehat{\pmb{x}}_{{m,k},{(t)}}\) can be modeled as signal $\pmb{x}_{m,k}$ plus independent Gaussian noise \(\pmb{v}_{m,k}\sim \mathcal{CN}(\textbf{0},{\pmb I}_N)\), i.e.,
\begin{align}\label{noise_model}
\widehat{\pmb{x}}_{{m,k},{(t)}}=\pmb{x}_{m,k}+\pmb{\Theta}_{m,{(t)}}^{\frac{1}{2}}\pmb{v}_{m,k},    
\end{align}
where \(\pmb{v}_{m,k}\) is independent of \(\pmb{x}_{m,k}\), and the state $\pmb{\Theta}_{m,{(t)}}$ can be predicted according to the state evolution. Starting from the initial state 
\begin{align}\label{state evolution_initial}
\pmb{\Theta}_{m,{(0)}}=\frac{\pmb{I}}{E}+
	\frac{|\mathcal{D}_m|}{\tau} \mathbb{E}_{\mathcal{D}_m}\bigg[\pmb{X}_{m,k}\pmb{X}_{m,k}^H\bigg],
\end{align}
the state evolution of the AMP algorithm at each iteration is updated as \cite{L. Liu_1}, i.e.,
\begin{flalign}\label{eqn:state evolution}
\pmb{\Theta}_{m,{(t+1)}}=\frac{\pmb{I}}{E}+
	&\frac{|\mathcal{D}_m|}{\tau} \mathbb{E}_{\mathcal{D}_m}\left\{\left[\eta(\widehat{\pmb{X}}_{{m,k},{(t)}})-\pmb{X}_{m,k}\right]\right.\nonumber\\
&\qquad \qquad 	\left.\left[\eta(\widehat{\pmb{X}}_{{m,k},{(t)}})-\pmb{X}_{m,k}\right]^H\right\},
\end{flalign}
where \(\pmb{X}_{m,k}\), \(\pmb{V}_{m,k}\) and  \(\widehat{\pmb{X}}_{{m,k},{(t)}}=\pmb{X}_{m,k}+\pmb{\Theta}_{m,{(t)}}^{\frac{1}{2}}\pmb{V}_{m,k}\) are used to capture the distribution of \(\pmb{x}_{m,k}\), \(\pmb{v}_{m,k}\), and $\widehat{\pmb{x}}_{{m,k},{(t)}}$, respectively. The expectation is taken over all of the devices in $\mathcal{D}_m$.

For a given \(\widehat{\pmb{x}}_{{m,k},{(t)}}\), the MMSE denoiser \(\eta(\cdot)\) is designed the same as in \cite{L. Liu_1}, given by
\begin{align}
 &\eta(\widehat{\pmb{x}}_{{m,k},{(t)}})=\mathbb{E}\left[{{\pmb X }}_{m,k}|\widehat{{\pmb X }}_{{m,k},{(t)}}
 =\widehat{{\pmb x }}_{{m,k},{(t)}}\right]\nonumber\\&=\phi_{{m,k},{(t)}}\beta_{m,k}\left(\beta_{m,k}{\pmb I}+{\pmb \Theta}_{m,{(t)}}\right)^{-1}\widehat{{\pmb x }}_{{m,k},{(t)}},\label{MMSE_denoiser_def}
\end{align}
where
\begin{align}
& \phi_{{m,k},{(t)}}=\frac{1}{{1+\frac{1-\epsilon}{\epsilon}{\rm exp}\left[-N(\mu_{{m,k},{(t)}}-\nu_{{m,k},{(t)}})\right]}},\tag{\ref{MMSE_denoiser_def}\text{a}}\\
&\mu_{{m,k},{(t)}}=\frac1{N}{{\widehat{{\pmb x }}_{{m,k},{(t)}}^H  \left\{{\pmb \Theta}_{m,{(t)}}^{-1}-[\beta_{m,k}{\pmb I}+{\pmb \Theta}_{m,{(t)}}]^{-1}\right\}\widehat{{\pmb x }}_{{m,k},{(t)}}}}, \tag{\ref{MMSE_denoiser_def}\text{b}}\\
& \nu_{{m,k},{(t)}}=\frac{1}{N}\log \det\left({\pmb I}+{\beta_{m,k}}{\pmb \Theta}_{m,{(t)}}^{-1}\right).\tag{\ref{MMSE_denoiser_def}\text{c}}
\end{align}

\begin{theorem}
In the asymptotic regime, the state evolution given in (\ref{eqn:state evolution}) with the MMSE denoiser (\ref{MMSE_denoiser_def}) can be simplified as
\begin{align}\label{state_eve_simple}
    {\pmb \Theta}_{m,{(t)}}=\theta_{m,{(t)}}{\pmb I},
\end{align}
with 
\begin{align}
& \theta_{m,{(0)}}=\frac{1}{E}+\frac{\epsilon |\mathcal{D}_m|}{\tau}  \mathbb{E}_{\mathcal{D}_m}[\beta_{m,k}],\tag{\ref{state_eve_simple}\text{a}}
\end{align}
and
\begin{flalign}
 \theta_{m,{(t+1)}}=\frac{1}{E}&+
	\frac{|\mathcal{D}_m|}{N\tau} \mathbb{E}_{\mathcal{D}_m}\left\{\left({\xi}_{{m,k},{(t)}}-1\right)^2\left\|{{\pmb X }}_{m,k}\right\|^2\right.\nonumber\\
	&\left.+\left(\xi_{{m,k},{(t)}}\right)^2\theta_{m,{(t)}}\left\|{{\pmb V }}_{m,k}\right\|^2\right\}, \tag{\ref{state_eve_simple}\text{b}}
\end{flalign}
where the expectation is taken over all of the devices in $\mathcal{D}_m$, and 
\begin{align}
    \xi_{{m,k},{(t)}}=\frac{\phi_{{m,k},{(t)}}\beta_{m,k}}{\beta_{m,k}+\theta_{m,{(t)}}}.\tag{\ref{state_eve_simple}\text{c}}
\end{align}

\end{theorem}

\begin{IEEEproof}
See Appendix C.
\end{IEEEproof}
Since \({\pmb \Theta}_{m,(t)}\) is a weighted identity matrix in each iteration, by substituting (\ref{state_eve_simple}) into (\ref{MMSE_denoiser_def}), the MMSE denoiser can be simplified as
\begin{align}\label{Simple_denoiser}
\eta(\widehat{{\pmb x }}_{{m,k},{(t)}})
& =\xi_{{m,k},{(t)}}\widehat{{\pmb x }}_{{m,k},{(t)}}, ~~~ \forall t,m,k,
\end{align}
where $\phi_{{m,k},{(t)}}$ in $\xi_{m,k}^{(t)}$ is defined in (\ref{MMSE_denoiser_def}) with
\begin{align}
&\mu_{{m,k},{(t)}}=\frac1{N}\left(\frac{1}{\theta_{m,{(t)}}}-\frac{1}{\beta_{m,k}+\theta_{m,{(t)}}}\right){||\widehat{{\pmb x }}_{{m,k},{(t)}}||^2}, \tag{\ref{Simple_denoiser}\text{a}}\\
& \nu_{{m,k},{(t)}}=\log\left(1+\frac{\beta_{m,k}}{\theta_{m,{(t)}}}\right). \tag{\ref{Simple_denoiser}\text{b}}
\end{align}

\subsection{Joint User Activity Detection for Stage I}
After the $t$-th iteration, the $m$-th AP obtains the output \(\{\widehat{\pmb{x}}_{{m,k},{(t)}},k\in \mathcal{D}_m\}\). According to the model (\ref{noise_model}), the entities of \(\widehat{\pmb{x}}_{{m,k},{(t)}}\) are i.i.d. zero-mean complex Gaussian variables with variances \(\beta_{m,k}+\theta_{m,{(t)}} \) and \(\theta_{m,{(t)}} \), and the corresponding PDFs  $f_1(\widehat{\pmb{x}}_{{m,k},{(t)}})$ and $f_0(\widehat{\pmb{x}}_{{m,k},{(t)}})$, for $\alpha_k=1$ and $\alpha_k=0$, respectively. Similar to the cellular massive MIMO \cite{L. Liu_1}, each AP can separately solve the following hypothesis testing problem 
\begin{align}\label{seperate_test}
\left\{\begin{array}{ll}\mathcal{H}_0: & \alpha_{k}=0, \\
\mathcal{H}_1: & \alpha_{k}=1,
\end{array}\right. 
\end{align}
whose optimal decision rule is the log-likelihood ratio given by
\begin{align}
\frac{{\rm Pr}(\alpha_k=1|\widehat{\pmb{X}}_{{m,k},{(t)}}=\widehat{\pmb{x}}_{{m,k},{(t)}})}{{\rm Pr}(\alpha_k=0|\widehat{\pmb{X}}_{{m,k},{(t)}}=\widehat{\pmb{x}}_{{m,k},{(t)}})}=\frac{f_1(\widehat{\pmb{x}}_{{m,k},{(t)}})}{f_0(\widehat{\pmb{x}}_{{m,k},{(t)}})}\mathop {\lessgtr }\limits _{\mathcal{H}_1}^{\mathcal{H}_0} 1,
\end{align}
which can be further simplified as 
\begin{align}
    \mu_{{m,k},{(t)}}\mathop {\lessgtr }\limits _{\mathcal{H}_1}^{\mathcal{H}_0} \nu_{{m,k},{(t)}}.
\end{align}
To reap the benefits from multiple APs and the user-centric architecture, the distributed APs, which are denoted by $\widetilde{\mathcal{A}}_k$ within a circular area with radius $R_1\leq R_0$, jointly detect the activity \({\widehat \alpha}_k\). Each AP $m\in \mathcal{A}_k$ sends \(\mu_{m,k,{(t)}}\) to the CPU, which determines \({\widehat \alpha}_k\) according to the following weighted rule,
\begin{align}\label{joint_judge_rule}
    {\widehat \alpha}_k=\left\{\begin{array}{ll}1, & {\rm if} ~
  \mu_{k,{(t)}}\geq \nu_{k,{(t)}}, \\
0, & {\rm if} ~
\mu_{k,{(t)}} < \nu_{k,{(t)}},
\end{array}\right. 
\end{align}
where \begin{align}
    \mu_{k,{(t)}}&=\sum\nolimits_{m\in \widetilde{\mathcal{A}}_k}\zeta_{m,k}\mu_{m,k,{(t)}},\tag{\ref{joint_judge_rule}\text{a}}\\
    \nu_{k,{(t)}}&=\sum\nolimits_{m\in \widetilde{\mathcal{A}}_k}\zeta_{m,k}\nu_{m,k,{(t)}},\tag{\ref{joint_judge_rule}\text{b}}
\end{align}
with the fusing weights ${\pmb \zeta}_k=(\cdots,\zeta_{m,k},\cdots), m\in \widetilde{\mathcal{A}}_k$ satisfying $\sum\nolimits_{m\in \widetilde{\mathcal{A}}_k}\zeta_{m,k}=1$.

Then, the CPU sends the estimated activity \({\widehat \alpha}_k\) to the APs $m\in \mathcal{A}_k$ within the coverage area of the $k$-th device for the following channel re-estimation. It is noted that the backhaul overhead can be neglected, because each AP only needs to deliver a scalar $\mu_{m,k,{(t)}}$ to the CPU.

\subsection{LMMSE Channel Re-estimation for Stage II}
The LMMSE channel estimation in cell-free systems have been already considered in \cite{S_Rao,Xinhua Wang1}. After obtaining \(\widehat{\pmb \alpha}_m\), we re-estimate the channel \(\widehat{\pmb{g}}_{(m,n)}\) using the LMMSE method to further improve the accuracy of channel estimation. From (\ref{receivd_pilot_EAN}), we have 
\begin{align}
  &\mathbb{E}\left[ {\pmb{y}}_{(m,n)}{\pmb{y}}_{(m,n)}^H\right]=\pmb{Z}_m=E\pmb{\Psi}_m\pmb{U}_m\pmb{\Psi}_m^H+\pmb{I},\label{Y_mean_square}\\
 & \mathbb{E}\left[ {\pmb{g}}_{(m,n)}{\pmb{y}}_{(m,n)}^H\right]=\sqrt{E}{\pmb{U}}_m\pmb{\Psi}_m^H, \label{Y_mean_square1}
\end{align}
where \({\pmb{U}}_m=  {\rm diag} \left(\cdots, {\alpha}_{k}\beta_{m,k},\cdots\right), k\in \mathcal{D}_m\). Since only \(\widehat{\pmb \alpha}_m={\rm diag} \left(\cdots, \widehat{\alpha}_{k}\cdots\right), k\in \mathcal{D}_m\) and $\{\beta_{m,k}, k\in \mathcal{D}_m\}$ are available at the $m$-th AP in (\ref{Y_mean_square}) and (\ref{Y_mean_square1}), we use  \(\widehat{\pmb{U}}_m= {\rm diag} \left(\cdots,\widehat{\alpha}_{k}\beta_{m,k},\cdots\right), k\in \mathcal{D}_m\) instead of ${\pmb{U}}_m$ for the LMMSE channel estimation. Thus, the LMMSE estimate of \(\pmb{g}_{(m,n)} \) is 
	\begin{align}\label{channel_estimate}
\widehat{\pmb{g}}_{(m,n)}&=\mathbb{E}\left[ {\pmb{g}}_{(m,n)}{\pmb{y}}_{(m,n)}^H\right]\left(\mathbb{E}\left[ {\pmb{y}}_{(m,n)}{\pmb{y}}_{(m,n)}^H\right]\right)^{-1}\pmb{y}_{(m,n)}\nonumber\\
&=\sqrt{E}\widehat{\pmb{U}}_m\pmb{\Psi}_m^H\widehat{\pmb{Z}}_m ^{-1}\pmb{y}_{(m,n)},
\end{align}
where $\widehat{\pmb{Z}}_m=E\pmb{\Psi}_m\widehat{\pmb{U}}_m\pmb{\Psi}_m^H+\pmb{I}$. Since ${\pmb{U}}_m$ and $\widehat{\pmb{U}}_m$ are diagonal matrices, ${\pmb{Z}}_m$ and $\widehat{\pmb{Z}}_m$ can be rewritten as 
\begin{align}\label{Z_M}
 \pmb{Z}_m=E\sum\nolimits_{i\in {\Omega}_m}\beta_{m,i}\pmb{\psi}_{m,i}\pmb{\psi}_{m,i}^H+\pmb{I}, 
\end{align}
and 
\begin{align}\label{Z_M_est}
  \widehat{\pmb{Z}}_m=E\sum\nolimits_{i\in \widehat{\Omega}_m}\beta_{m,i}\pmb{\psi}_{m,i}\pmb{\psi}_{m,i}^H+\pmb{I},
\end{align}
where $\widehat{\Omega}_m=\{i:\widehat{\alpha}_{i}=1,i \in \mathcal{D}_m\}$. We can easily observe that
$$\Omega_m=\left(\widehat{\Omega}_m\cup\widehat{\Omega}_m^{\rm M}\right) \setminus \widehat{\Omega}_m^{\rm F}, $$
where $\widehat{\Omega}_m^{\rm M}=\{i:\alpha_{i}=1,\widehat{\alpha}_{i}=0, i \in \mathcal{D}_m\}$ is the set of miss detected devices in $\mathcal{D}_m$, and $\widehat{\Omega}_m^{\rm F}=\{i:\alpha_{i}=0,\widehat{\alpha}_{i}=1,i \in \mathcal{D}_m\}$ is the set of false detected devices in $\mathcal{D}_m$. 
\begin{table}[t!]
\renewcommand{\arraystretch}{1.3}
\caption{}
\label{table_parameters2} \vskip-2mm
\centering
\begin{tabular}{p{1cm}p{7cm}}
\hline
\bfseries Notation & \bfseries Meaning \\
 \hline $\Omega_m$ & Active devices within the coverage of the $m$-th AP with ${\Omega}_m=\{k:\alpha_k=1, d_{m,k}\leq R_0\}$\\
  \hline $\widehat{\Omega}_m$ & Estimated active devices within the coverage of the $m$-th AP with $\widehat{\Omega}_m=\{k:\widehat{\alpha}_k=1, d_{m,k}\leq R_0\}$ \\
    \hline $\widehat{\Omega}^M_m$ & Miss detected devices within the coverage of the $m$-th AP with $\widehat{\Omega}^M_m=\{k:{\alpha}_k=1, \widehat{\alpha}_k=0, d_{m,k}\leq R_0\}$\\
      \hline $\widehat{\Omega}^F_m$ & False detected devices within the coverage of the $m$-th AP with $\widehat{\Omega}^M_m=\{k:{\alpha}_k=0, \widehat{\alpha}_k=1, d_{m,k}\leq R_0\}$\\
\hline
\end{tabular}
\end{table}

Therefore, we have 
	\begin{align}\label{difference_Z_M}
	    {\pmb{Z}}_m&=\widehat{\pmb{Z}}_m+\widetilde{\pmb{Z}}_m,
	\end{align}
	where \begin{align}\label{B_M}
	    \widetilde{\pmb{Z}}_m=E\sum\nolimits_{{i}\in \widehat{\Omega}_m^{\rm M}}\beta_{m,{i}}\pmb{\psi}_{{i}}\pmb{\psi}_{{i}}^H-E\sum\nolimits_{{j}\in \widehat{\Omega}_m^{\rm F}}\beta_{m,{j}}\pmb{\psi}_{{j}}\pmb{\psi}_{{j}}^H.
	\end{align}
	Substituting (\ref{Y_mean_square}) and (\ref{difference_Z_M}) into (\ref{channel_estimate}), we can obtain 
		\begin{align}
	\mathbb{E}\left[ \widehat{\pmb{g}}_{(m,n)}\widehat{\pmb{g}}_{(m,n)}^H\right]
	&=E\widehat{\pmb{U}}_m\pmb{\Psi}_m^H\widehat{\pmb{Z}}_{m}^{-1}\pmb{\Psi}_m\widehat{\pmb{U}}_m\nonumber\\
	&+E\widehat{\pmb{U}}_m\pmb{\Psi}_m^H\widehat{\pmb{Z}}_{m}^{-1}\widetilde{\pmb{Z}}_m\widehat{\pmb{Z}}_{m}^{-1}\pmb{\Psi}_m\widehat{\pmb{U}}_m.\nonumber
	\end{align}
Then, the mean square of the channel estimate for the $k$-th device can be expressed as
	\begin{align}\label{mean_squre_channel_est}
	 \gamma_{m,k}&=\mathbb{E}\left[\left|{\widehat g}_{(m,n),k}\right|^2\right]=\mathbb{E}\left[ \widehat{\pmb{g}}_{(m,n)}\widehat{\pmb{g}}_{(m,n)}^H\right]_{kk}\nonumber\\
	 &=E\beta^2_{m,k}\pmb{\psi}_{k}^H\left[\widehat{\pmb{Z}}_m^{-1}+\widehat{\pmb{Z}}_{m}^{-1}\widetilde{\pmb{Z}}_m\widehat{\pmb{Z}}_{m}^{-1}\right]\pmb{\psi}_{k}.
	\end{align}
	and the corresponding mean-squared channel estimation error is 
	\begin{align}\label{mean_error}
    e_{m,k}=\mathbb{E}\left[\left\|{g}_{(m,n),k}-{\widehat g}_{(m,n),k}\right\|^2\right]=\beta_{m,k}-\gamma_{m,k}.
\end{align}
\emph{\textbf{Remark 1}}: Noted that the activity $\alpha_k$ is jointly detected by the adjacent APs, while the CSI is estimated at each AP using the LMMSE method. Since the CSI of different antennas is independent, the antennas' cooperation gains no benefits for the channel estimation when the active devices are known. It can be seen from \cite{ J. Jose} that, the optimal LMMSE channel estimation is performed at each antenna separately. In brief, the joint activity detection can improve the accuracy of activity detection and further improve the performance of channel estimation.


\section{Performance Analysis for Each device}
Next, we analyze the the activity detection error 
probability and the mean-squared error of channel estimation for the $k$-th device.
\subsection{Activity Detection Error Probability}
\begin{definition}
Given the fusing weights ${\pmb \zeta}_k$, the {\em activity detection error probability} for the $k$-th device is defined as
\begin{align}\label{errorpro}
    {\mathcal{P}}_k({\pmb \zeta}_k)=\epsilon{\mathcal{P}}^M_k({\pmb \zeta}_k)+(1-\epsilon){\mathcal{P}}^F_k({\pmb \zeta}_k),
\end{align}
where ${\mathcal{P}}^M_k({\pmb \zeta}_k)$ is {\em the probability of missed detection} given by
\begin{align}
    {\mathcal{P}}^M_k({\pmb \zeta}_k)={\rm Pr}({\widehat \alpha}_k=0|{\alpha}_k=1)={\rm Pr}(\mu_{k,(t)}<\nu_{k,(t)}|{\alpha}_k=1), \tag{\ref{errorpro}\text{a}}
\end{align}
and ${\mathcal{P}}^F_k({\pmb \zeta}_k)$ is {\em the probability of false detection} given by
\begin{align}
    {\mathcal{P}}^F_k({\pmb \zeta}_k)={\rm Pr}({\widehat \alpha}_k=1|{\alpha}_k=0)={\rm Pr}(\mu_{k,(t)}>\nu_{k,(t)}|{\alpha}_k=0).\tag{\ref{errorpro}\text{b}}
\end{align}
\end{definition}
From (\ref{noise_model}), it follows that 
\begin{align}\label{individual_dis}
 \zeta_{m,k}\mu_{m,k,{(t)}}\sim\left\lbrace\begin{aligned}
   &{\rm Gamma}(N,{\vartheta}_{m,k}) ,& {\rm if}~ \alpha_k=1,\\
   &{\rm Gamma}(N,{\vartheta}'_{m,k}),&  {\rm if} ~ \alpha_k=0,\\
   \end{aligned}
   \right. 
\end{align}
where $${\vartheta}_{m,k}={\zeta_{m,k}\beta_{m,k}}/{N\theta_{m,(t)}},$$ $${\vartheta}'_{m,k}={\zeta_{m,k}\beta_{m,k}}/{[N(\theta_{m,(t)}+\beta_{m,k})]}.$$
According to (\ref{joint_judge_rule}\text{a}), $\mu_{k,(t)}$ is a sum of independent Gamma random variables with different scales, which implies the PDF of $\mu_{k,(t)}$ is too complex to analyze. To facilitate analysis, we approximate the distribution of \(\mu_{k,(t)}\) by a Gamma distribution according to the well-known Welch-Satterthwaite approximation \cite{F. E. Satterthwaite, B. L. Welch}, i.e., $\mu_{k,(t)}\sim {\rm Gamma}(s,\omega)$, with
\begin{align}
 \mu_{k,(t)}\sim \left\lbrace\begin{aligned}
   &{\rm Gamma}(s_k,{\omega}_{k}) ,& {\rm if}~ \alpha_k=1,\\
   &{\rm Gamma}(s'_k,{\omega}'_{k}),&  {\rm if} ~ \alpha_k=0,\\
   \end{aligned}
   \right. 
\end{align}
where 
\begin{align}
    &s_k=\frac{N\left(\sum_{m\in \widetilde{\mathcal{A}}_k}\vartheta_{m,k}\right)^2}{\sum_{m\in \widetilde{\mathcal{A}}_k}\vartheta^2_{m,k}}, \omega_k=\frac{\sum_{m\in \widetilde{\mathcal{A}}_k}\vartheta^2_{m,k}}{\sum_{m\in \widetilde{\mathcal{A}}_k}\vartheta_{m,k}},\nonumber\\
    &s'_k=\frac{N\left(\sum_{m\in \widetilde{\mathcal{A}}_k}\vartheta'_{m,k}\right)^2}{\sum_{m\in \widetilde{\mathcal{A}}_k}(\vartheta'_{m,k})^2}, \omega'_k=\frac{\sum_{m\in \widetilde{\mathcal{A}}_k}(\vartheta_{m,k})^2}{\sum_{m\in \widetilde{\mathcal{A}}_k}\vartheta'_{m,k}}.\nonumber
\end{align}
According to the cumulative distribution function (CDF) of the Gamma distribution, the probability of missed detection \({\mathcal P}^M_k({\pmb \zeta}_k)\) can be approximated as 
\begin{align}\label{P_miss_app}
    \widetilde{\mathcal P}^M_k({\pmb \zeta}_k)={\rm Pr}(\mu_{k,(t)}<\nu_{k,(t)}|\alpha_k=1)=\frac{{\gamma}(s_k,{\nu_{k,(t)}}/{\omega_k})}{\Gamma(s_k)}.
\end{align}
 Similarly, the probability of false detection \({\mathcal P}^F_k({\pmb \zeta}_k)\) can be approximated as
\begin{align}\label{P_fal_app}
    \widetilde{\mathcal P}^F_k({\pmb \zeta}_k)&={\rm Pr}(\mu_{k,(t)}>\nu_{k,(t)}|\alpha_k=0)=\frac{{\Gamma}(s'_k,{\nu_{k,(t)}}/{\omega'_k})}{\Gamma(s'_k)}.
\end{align}
According to (\ref{errorpro}), (\ref{P_miss_app}), and (\ref{P_miss_app}), the activity detection error probability \({\mathcal P}_k({\pmb \zeta}_k)\) can be approximated as
\begin{align}\label{P_error_app}
    \widetilde{\mathcal P}_k({\pmb \zeta}_k)=\epsilon\frac{{\gamma}(s_k,{\nu_{k,(t)}}/{\omega_k})}{\Gamma(s_k)}+(1-\epsilon)\frac{{\Gamma}(s'_k,{\nu_{k,(t)}}/{\omega'_k})}{\Gamma(s'_k)}.
\end{align}
Simulation results will show that the error caused by the Welch-Satterthwaite approximation is negligible.

Since \(\widetilde{{\mathcal P}}_k({\pmb \zeta}_k)\) is an accurate approximation of \({\mathcal P}_k({\pmb \zeta}_k)\), the optimal fusing weights for the $k$-th device can be expressed as 
\begin{align}\label{opt_fusion}
    {\pmb \zeta}^{*}_k=\mathop{\arg \min} {\mathcal P}_k({\pmb \zeta}_k),
\end{align}
and it can be exhaustively searched through minimizing  $\widetilde{\mathcal P}_k({\pmb \zeta}_k)$ which depends on the distribution of $\{\beta_{m,k},m\in \widetilde{\mathcal{A}}_k\}$. 

\emph{\textbf{Remark 2}}: The small cell system can be treated as a special case of the cell-free IoT, in which the activity of the $k$-th device $\alpha_k$ is only determined by the nearest AP with index $m^o$, i.e., $\beta_{m^o,k}=\max_{m\in \mathcal{A}_k}\beta_{m,k}$. That is, only $\zeta'_{m^o,k}=1$ in the fusing weight is non-zero for the small cell system, i.e.,
\begin{align}\label{sc_fusion}
    {\pmb \zeta}'_k=(\zeta'_{m^o,k}=1,\zeta'_{m,k}=0,\cdots), m\neq m^o,~m\in \widetilde{\mathcal{A}}_k.
\end{align}

\subsection{Impacts of $N$ and $\lambda_A$}

Intuitively, the number of antennas $N$ of each AP and the density of APs $\lambda_A$ play important rules on the accuracy of activity detection. To show this, we first introduce the following Lemma.
\begin{lemma}
Given \(c\in (0,1)\) and \(c'\in (1,+\infty)\), we have 
\begin{align}
    \mathop{\lim}_{{s}\rightarrow +\infty} {{\gamma}({s},c{s})}/{\Gamma({s})}=0,\label{sinf1}\\
    \mathop{\lim}_{{s}\rightarrow +\infty} {{\Gamma}({s},c'{s})}/{\Gamma({s})}=0.\label{sinf2}
\end{align}
\end{lemma}
\begin{IEEEproof}
See Appendix B.
\end{IEEEproof}
According to (\ref{P_miss_app}), (\ref{P_fal_app}) and Lemma 1, we have the following theorem. 
\begin{theorem}
With \(N\rightarrow \infty\), the activity detection error probability with optimal fusing weight ${\pmb \zeta}^{*}_k$ approaches zero, i.e., $$\mathop{\lim}_{N\rightarrow\infty}\widetilde{\mathcal P}_k({\pmb \zeta}^{*}_k)=0.$$
\end{theorem}
\begin{IEEEproof} 
	It is sufficient for us to consider a small-cell system. We will adopt notations used in Remark 2. According to (\ref{opt_fusion}), we have 
\begin{align}\label{P_k_opt}
\widetilde{\mathcal P}_k({\pmb \zeta}^{*}_k)\leq \widetilde{\mathcal P}_k({\pmb \zeta}'_k)=\epsilon\widetilde{\mathcal P}_k^M({\pmb \zeta}'_k)+(1-\epsilon)\widetilde{\mathcal P}_k^F({\pmb \zeta}'_k).
\end{align}
Since $\mu_{k,(t)}({\pmb \zeta}'_k)=\mu_{m^o,k,{(t)}}$ follows a Gamma distribution and $$\nu_{k,(t)}({\pmb \zeta}'_k)=\nu_{m^o,k,{(t)}}=\log\left(1+\beta_{m^o,k}/\theta_{m^o,(t)}\right),$$ we have 
 \begin{align}
     \widetilde{\mathcal P}_k^M({\pmb \zeta}'_k)={\gamma(N,cN)}/{\Gamma(N)}, \\
     \widetilde{\mathcal P}_k^F({\pmb \zeta}'_k)={\Gamma(N,c'N)}/{\Gamma(N)},
 \end{align}
with \begin{align}
    c&=\frac{\log\left(1+{\beta_{m^o,k}}/{\theta_{m^o,(t)}}\right)}{{\beta_{m^o,k}}/{\theta_{m^o,(t)}}}\overset{(a)}{<}1,\nonumber\\
    c'&=\frac{\log\left(1+{\beta_{m^o,k}}/{\theta_{m^o,(t)}}\right)}{{\beta_{m^o,k}}/{(\theta_{m^o,(t)}+\beta_{m^o,k}})}\overset{(b)}{>}1,\nonumber
\end{align}
where steps (a) and (b) follow from $\frac{x}{1+x}<\log(1+x)<x$ for $x>0$. Using Lemma 1, we have $\mathop{\lim}_{N\rightarrow\infty}\widetilde{\mathcal P}_k^M({\pmb \zeta}'_k)=0$ and $\mathop{\lim}_{N\rightarrow\infty}\widetilde{\mathcal P}_k^F({\pmb \zeta}'_k)=0$. Taking the limits on both sides of (\ref{P_k_opt}), we conclude the proof.
\end{IEEEproof}

\begin{theorem}
	Define the equal fusing weights as
	\begin{align}\label{equ_fusing}
	{\pmb \zeta}^{\#}_k=\left({1}/{|\widetilde{\mathcal{A}}_k|},\cdots,{1}/{|\widetilde{\mathcal{A}}_k|}\right).
	\end{align}
	As \(|\widetilde{\mathcal{A}}_k|\rightarrow \infty\), the activity detection error probability with fusing weight ${\pmb \zeta}^{*}_k$ and ${\pmb \zeta}^{\#}_k$ approaches 0, i.e., $${\mathop{\lim}}_{|\widetilde{\mathcal{A}}_k|\rightarrow\infty} \widetilde{\mathcal P}_k({\pmb \zeta}^{*}_k)={\mathop{\lim}}_{|\widetilde{\mathcal{A}}_k|\rightarrow\infty} \widetilde{\mathcal P}_k({\pmb \zeta}^{\#}_k)=0.$$
\end{theorem}
\begin{IEEEproof} 
	We have 
	\begin{align}\label{P_k_opt1}
	\widetilde{\mathcal P}_k({\pmb \zeta}^{*}_k)\leq \widetilde{\mathcal P}_k({\pmb \zeta}^{\#}_k)=\widetilde{\mathcal P}_k^M({\pmb \zeta}^{\#}_k)+\widetilde{\mathcal P}_k^F({\pmb \zeta}^{\#}_k).
	\end{align}
	
	As $\alpha_k=1$, we have $\mu_{m,k,{(t)}}\sim {\rm Gamma}(N,{\vartheta}_{m,k})$ with ${\vartheta}_{m,k}={\beta_{m,k}}/{N\theta_{m,(t)}}$. Introducing auxiliary variables ${\mu}'_{m,k}\sim {\rm Gamma}(N,{\vartheta}_{\min})$ and ${\nu}'_{m,k}=\log(1+N\widetilde{\vartheta}_{\min})$
	with ${\vartheta}_{\min}={\beta_{\min}}/{(N\theta_{\max})}\leq {\vartheta}_{m,k}$,  $\beta_{\min}=\min\nolimits_{m\in \widetilde{\mathcal{A}}_k}\beta_{m,k},~\mbox{and}~\theta_{\max}=\max\nolimits_{m\in \widetilde{\mathcal{A}}_k}\theta_{m,(t)}.$ For any $\Delta,$ we have
	\begin{align} 
	&{\rm Pr}({\mu}_{m,k,(t)}\leq {\nu}_{m,k,(t)}+\Delta
	|{\alpha}_k=1,\nu-\mu=\Delta)\nonumber\\
	&={\gamma\left(N,\left[\log\left(1+N{\vartheta}_{m,k}\right)+\Delta\right]/{\vartheta}_{m,k}\right)}/{\Gamma(N)}\nonumber\\
	&{\leq} {\rm Pr}(\widetilde{\mu}'_{m,k}\leq \widetilde{\nu}'_{m,k}+\Delta
	|{\alpha}_k=1,\nu-\mu=\Delta) \nonumber\\
	&={\gamma\left(N,\left[\log\left(1+N{\vartheta}_{\min}\right)+\Delta\right]/{\vartheta}_{\min}\right)}/{\Gamma(N)},\nonumber
	\end{align}
	where the inequality follows from the monotonicity of $f(x)={\gamma\left(N,\left[\log\left(1+Nx\right)+\Delta\right]/x\right)}/{\Gamma(N)}$. Thus, we have 
	\begin{align}\label{propos1.2}
	&{\rm Pr}({\mu}+\mu_{m,k,(t)}\leq {\nu}+\nu_{m,k,(t)}|{\alpha}_k=1),\nonumber\\
	& \leq {\rm Pr}({\mu}+\widetilde{\mu}'_{m,k}\leq {\nu}+\widetilde{\nu}'_{m,k}|{\alpha}_k=1).
	\end{align}
	Using (\ref{propos1.2}) repeatedly, we obtain 
	\begin{align}
	\widetilde{\mathcal P}^M_k({\pmb \zeta}^{\#}_k)&={\rm Pr}(\sum\nolimits_{m\in \widetilde{\mathcal{A}}_k}\mu_{m,k,(t)}<\sum\nolimits_{m\in \widetilde{\mathcal{A}}_k}\nu_{m,k,(t)}|\alpha_k=1) \nonumber\\
	&\leq {\rm Pr}(\sum\nolimits_{m\in \widetilde{\mathcal{A}}_k}{\mu}'_{m,k}<\sum\nolimits_{m\in \widetilde{\mathcal{A}}_k}{\nu}'_{m,k}|\alpha_k=1)\nonumber\\
	&=\frac{\gamma\left(N|\widetilde{\mathcal{A}}_k|,|\widetilde{\mathcal{A}}_k|\log\left(1+N{\vartheta}_{\min}\right)/{\vartheta}_{\min}\right)}{\Gamma(N|\widetilde{\mathcal{A}}_k|)}.
	\end{align}
	Using Lemma 1, we have $\mathop{\lim}_{|\widetilde{\mathcal{A}}_k|\rightarrow \infty}\widetilde{\mathcal P}_k^M({\pmb \zeta}^{\#}_k)=0$. Using the similar arguments, we obtain $\mathop{\lim}_{|\widetilde{\mathcal{A}}_k|\rightarrow \infty}\widetilde{\mathcal P}_k^F({\pmb \zeta}^{\#}_k)=0$. Taking the limit in both side of (\ref{P_k_opt1}), we can conclude the proof.
\end{IEEEproof}

\subsection{Mean Squared Channel Estimation Error }
 Let $\mathcal{X}$ be an arbitrary subset of $\mathcal{D}_m$, i.e., $\mathcal{X}\subset \mathcal{D}_m $ and define \begin{align}\label{ZX}
    {\pmb Z}{(\mathcal{X})}&={\pmb I}+\sum_{i\in \mathcal{X} }E\beta_{m,i}\pmb{\psi}_{i}\pmb{\psi}_{i}^H\nonumber\\
    &={\pmb I}+\pmb{\Xi}(\mathcal{X})[\pmb{\Xi}(\mathcal{X})]^H,
\end{align} 
where \({\pmb \Xi}(\mathcal{X})=\pmb{\Psi}{(\mathcal{X})}[\pmb{\Pi}(\mathcal{X})]^{\frac1{2}}\), \(\pmb{\Psi}{(\mathcal{X})}=\left(\cdots,\pmb{\psi}_{m,i},\cdots\right)\), and \(\pmb{\Pi}(\mathcal{X})={\rm diag}\left(\cdots,E\beta_{m,i},\cdots\right)\) with $i\in \mathcal{X}$. Using Theorem 1 and Theorem 2 of \cite{J. Hoydis}, we can derive the following lemma.

\begin{lemma}
As \(\tau \rightarrow \infty\) keeping \(E\) fixed, we have 
\begin{align}\label{Lemma2_a}
    {{\rm tr}\left([\pmb{Z}{(\mathcal{X})}]^{-1}\right)}/{\tau}-{\mathcal{Q}{(\mathcal{X})}}\xrightarrow[]{\text{a.s.}}0,
\end{align}
and
\begin{align}\label{Lemma2_b}
    {{\rm tr}\left([\pmb{Z}{(\mathcal{X})}]^{-2}\right)}/{\tau}-{\overline{\mathcal{Q}}{(\mathcal{X})}}\xrightarrow[]{\text{a.s.}} 0,
\end{align}
where \(\mathcal{Q}{(\mathcal{X})}\) is given by
\begin{align}
  \mathcal{Q}{(\mathcal{X})}=\left(\sum\nolimits_{{i\in \mathcal{X}}}\frac{E\beta_{m,i}}{\tau(1+\varsigma_{i}) } +1 \right)^{-1},  
\end{align}
and \(\pmb{\varsigma}=\left[\cdots,\varsigma_{i},\cdots\right]^T,i\in\mathcal{X}\), is the unique solution of the following \(|\mathcal{X}|-1\) fixed-point equations
\[\varsigma_{i}^{(t+1)}={E\beta_{m,i}}\left(\sum\nolimits_{{i\in \mathcal{X}}}\frac{E\beta_{m,i}}{\tau(1+\varsigma^{(t)}_{i})} +1 \right)^{-1},\]
with initial values \(\varsigma_{i}=1\).
\(\overline{\mathcal{Q}}{(\mathcal{X})}\) is given by
\begin{align}
  \overline{\mathcal{Q}}{(\mathcal{X})}=&\left[1+\frac{1}{\tau}\sum\nolimits_{i\in \mathcal{X}}\frac{E\beta_{m,i}\widehat{\varsigma}_{i}}{(1+\varsigma_{i})^2 }\right][{\mathcal{Q}}{(\mathcal{X})}]^{2},  \nonumber
\end{align}
and \(\widehat{\pmb{\varsigma}}=\left[\cdots,\widehat{\varsigma}_{i},\cdots\right]^T,i\in \mathcal{X}_m\), is given by
\begin{flalign}
&\pmb{\varsigma}'=({\pmb I}-{\pmb J})^{-1}{\pmb r},\\
&[{\pmb J}]_{i,j}=\frac{E^2\beta_{m,i}\beta_{m,j}}{\tau(1+\varsigma_{i})^2 }[{\mathcal{Q}}{(\mathcal{X})}]^{2}, \\
&[{\pmb r}]_i=E\beta_{m,i}[{\mathcal{Q}}{(\mathcal{X})}]^{2}.
\end{flalign}
\end{lemma}

According to (\ref{mean_squre_channel_est}), $\gamma_{m,k}$ can be re-written as
\begin{align}\label{imperfect_gamma}
    \gamma_{m,k}=\widehat{\gamma}_{m,k}+\widetilde{\gamma}_{m,k},
\end{align}
where 
\begin{align}\label{gamma_1}\widehat{\gamma}_{m,k}=E\beta^2_{m,k}\pmb{\psi}_{k}^H\widehat{\pmb{Z}}_m^{-1}\pmb{\psi}_{k},\end{align} and \begin{align}\label{gamma_2}\widetilde{\gamma}_{m,k}=E\beta^2_{m,k}\pmb{\psi}_{k}^H\left[\widehat{\pmb{Z}}_{m}^{-1}\widetilde{\pmb{Z}}_m\widehat{\pmb{Z}}_{m}^{-1}\right]\pmb{\psi}_{k}.\end{align}

 \begin{theorem}
 	For \(\tau\rightarrow \infty\) with ${\tau}/{(\epsilon |\mathcal{D}_m|)}\leq \infty$ and $\rho={E}/{\tau}\rightarrow 0$, we have
 	\begin{align}\label{channel_app}
 	{e}_{m,k}-{\overline e}_{m,k}\xrightarrow[]{\text{a.s.}}0,
 	\end{align}
 	where ${e}_{m,k}$ is the mean squared channel estimation error defined in (\ref{mean_error}), and
 	${\overline e}_{m,k}$ is given by (\ref{imperfect_error}) at the top of next page.
 	\begin{figure*}
 		\begin{align}\label{imperfect_error}
 		{\overline e}_{m,k}=\frac{\beta_{m,k}}{1+{E}\beta_{m,k}\mathcal{Q}{(\widehat{\Omega}_m/\{k\})}}&-\sum_{i\in \widehat{\Omega}_m^{\rm M}}\frac{E^2\beta^2_{m,k}\beta_{m,i}{\overline{\mathcal{Q}}}{(\widehat{\Omega}_m/\{k\})}}{\tau[1+{E}\beta_{m,k}\mathcal{Q}{(\widehat{\Omega}_m/\{k\})}]^2 }\nonumber\\
 		& +\sum_{j\in \widehat{\Omega}_m^{\rm F}}\frac{E^2\beta^2_{m,k}\beta_{m,j}\overline{\mathcal Q}{(\widehat{\Omega}_m/\{k,j\})}}{\tau[1+{E}\beta_{m,k}\mathcal{Q}{(\widehat{\Omega}_m/\{k\})}]^2[1+{E}\beta_{m,j}\mathcal{Q}{(\widehat{\Omega}_m/\{k,j\})}]^2}\qquad \qquad  
 		\end{align}
 	\end{figure*}
 \end{theorem}
 \begin{IEEEproof}  
 	Using (\ref{Z_M_est}) and (\ref{ZX}), we have
 	\[\widehat{\pmb{Z}}_m={\pmb{Z}}{(\widehat{\Omega}_{m-\{k\}})}+E\beta_{m,k}\pmb{\psi}_{m,k}\pmb{\psi}_{m,k}^H.\] 
 	Substituting this into (\ref{gamma_1}), \(\widehat{\gamma}_{m,k}\) can be written as 
 	\begin{flalign}\label{gamma_matrix_invers}
 	\widehat{\gamma}_{m,k}
 	&=E\beta^2_{m,k}\pmb{\psi}_{k}^H\left({\pmb{Z}}{(\widehat{\Omega}_{m-\{k\}})}+E\beta_{m,k}\pmb{\psi}_{k}\pmb{\psi}_{k}^H\right)^{-1}\pmb{\psi}_{m,k}\nonumber\\
 	&=\frac{E\beta^2_{m,k}\pmb{\psi}_{k}^H[{\pmb{Z}}{(\widehat{\Omega}_{m-\{k\}})}]^{-1}\pmb{\psi}_{k}}{1+E\beta_{m,k}\pmb{\psi}_{k}^H[{\pmb{Z}}{(\widehat{\Omega}_{m-\{k\}})}]^{-1}\pmb{\psi}_{k}},
 	\end{flalign}
 	where the last step follows from (\ref{matrix_inverse_1}). According to (\ref{xAx}), we obtain
 	\begin{align}\label{gamma_hat}
 	\widehat{\gamma}_{m,k}-\frac{{E}\beta^2_{m,k}\mathcal{Q}{(\widehat{\Omega}_{m-\{k\}})}}{1+{E}\beta_{m,k}\mathcal{Q}{(\widehat{\Omega}_{m-\{k\}})}}\xrightarrow[]{\text{a.s.}}0.
 	\end{align}
 	Now we will estimate $\widetilde{\gamma}_{m,k}$. Substituting (\ref{B_M}) into (\ref{gamma_2}), we have
 	\begin{align}
 	\widetilde{\gamma}_{m,k}&=\sum_{i\in \widehat{\Omega}_m^{\rm M}}\underbrace{E^2\beta^2_{m,k}\beta_{m,i}\pmb{\psi}_{k}^H\widehat{\pmb{Z}}_{m}^{-1}\pmb{\psi}_{i}\pmb{\psi}_{i}^H\widehat{\pmb{Z}}_{m}^{-1}\pmb{\psi}_{k}}_{\mathcal{T}_{1}}\nonumber\\
 	&-\sum_{j\in \widehat{\Omega}_m^{\rm F}}\underbrace{E^2\beta^2_{m,k}\beta_{m,j}\pmb{\psi}_{k}^H\widehat{\pmb{Z}}_{m}^{-1}\pmb{\psi}_{j}\pmb{\psi}_{j}^H\widehat{\pmb{Z}}_{m}^{-1}\pmb{\psi}_{k}}_{\mathcal{T}_{2}}.
 	\end{align}
 	The first item $\mathcal{T}_1$ in $\widetilde{\gamma}_{m,k}$ can be further written as (\ref{gamma_miss}) at the top of next page. 
 	\begin{figure*}
 		\begin{align}\label{gamma_miss}
 		\mathcal{T}_1&\overset{(a)}{=}\frac{E^2\beta^2_{m,k}\beta_{m,i}\pmb{\psi}_{k}^H[{\pmb{Z}}{(\widehat{\Omega}_m/\{k\})}]^{-1}\pmb{\psi}_{i}\pmb{\psi}_{i}^H[{\pmb{Z}}{(\widehat{\Omega}_m/\{k\})}]^{-1}\pmb{\psi}_{k}}{\left(1+E\beta_{m,k}\pmb{\psi}_{k}^H[{\pmb{Z}}{(\widehat{\Omega}_m/\{k\})}]^{-1}\pmb{\psi}_{k}\right)^2}\overset{(b)}{=}\frac{E^2\beta^2_{m,k}\beta_{m,i}\pmb{\psi}_{i}^H[{\pmb{Z}}{(\widehat{\Omega}_m/\{k\})}]^{-2}\pmb{\psi}_{i}}{\tau\left[1+E\beta_{m,k} \mathcal{Q}{(\widehat{\Omega}_m/\{k\})} \right]^2}\nonumber\\
 		&\overset{(c)}{=}\frac{E^2\beta^2_{m,k}\beta_{m,i}\overline{\mathcal{Q}}{(\widehat{\Omega}_m/\{k\})}}{\tau\left[1+E\beta_{m,k} \mathcal{Q}{(\widehat{\Omega}_m/\{k\})} \right]^2}.
 		\end{align}
 		\hrulefill 
 	\end{figure*}
 	Step (a) follows from (\ref{matrix_inverse_1}), and step (b) follows from (\ref{xAx}) with $${\pmb A}=[\widehat{\pmb{Z}}{(\widehat{\Omega}_m/\{k\})}]^{-1}\pmb{\psi}_{i}\pmb{\psi}_{i}^H[\widehat{\pmb{Z}}{(\widehat{\Omega}_m/\{k\})}]^{-1}$$ for numerator, and step (c) follows from  (\ref{Lemma2_b}) and (\ref{xAx}). 
 	
 	The second item $\mathcal{T}_2$ in $\widetilde{\gamma}_{m,k}$, that is for the false detected device, with $j\in \widehat{\Omega}_m^{\rm F}$ and $j \in \widehat{\Omega}_m $, can be written as (\ref{gamma_false}) shown at the top of next page, where step (a) follows from (\ref{matrix_inverse_1}) and (\ref{matrix_inverse_2}) since $${\pmb{Z}}{(\widehat{\Omega}_m/\{k,j\})}={\pmb{Z}}{(\widehat{\Omega}_m/\{k\})}-{E}\beta_{m,j}\pmb{\psi}_{m,j}\pmb{\psi}_{m,j}^H.$$
 	Substituting (\ref{gamma_hat}), (\ref{gamma_miss}), and (\ref{gamma_false}) into (\ref{imperfect_gamma}), we conclude the proof.
 	\begin{figure*}
 		\begin{align}\label{gamma_false}
 		\mathcal{T}_2&{=}\frac{E^2\beta^2_{m,k}\beta_{m,j}\pmb{\psi}_{j}^H[{\pmb{Z}}{(\widehat{\Omega}_m/\{k\})}]^{-2}\pmb{\psi}_{j}}{\tau\left(1+E\beta_{m,k} \mathcal{Q}{(\widehat{\Omega}_m/\{k\})} \right)^2}\overset{(a)}{=}\frac{E^2\beta^2_{m,k}\beta_{m,j}\pmb{\psi}_{j}^H[{\pmb{Z}}{(\widehat{\Omega}_m/\{k,j\})}]^{-2}\pmb{\psi}_{j}}{\tau\left[1+E\beta_{m,k} \mathcal{Q}{(\widehat{\Omega}_m/\{k\})} \right]^2\left[1+{E}\beta_{m,j}\mathcal{Q}{(\widehat{\Omega}_m/\{k,j\})}\right]^2}\nonumber\\
 		&=\frac{E^2\beta^2_{m,k}\beta_{m,j}\overline{\mathcal{Q}}{(\widehat{\Omega}_m/\{k,j\})}}{\tau\left[1+E\beta_{m,k} \mathcal{Q}{(\widehat{\Omega}_m/\{k\})} \right]^2\left[1+{E}\beta_{m,j}\mathcal{Q}{(\widehat{\Omega}_m/\{k,j\})}\right]^2}.
 		\end{align}
 		\hrulefill 
 	\end{figure*}

 \end{IEEEproof}

\section{Performance Analysis for Network}
From (\ref{errorpro}), the activity detection error probability $\widetilde{\mathcal{P}}_k({\pmb \zeta}_k)$ is a function of the path-loss coefficients $\beta_{m,k}, m\in \widetilde{\mathcal{A}}_k$, which depends on the  geometry distribution of APs in $\widetilde{\mathcal{A}}_k$. 

The number of APs in $\widetilde{\mathcal{A}}_k$ is a random number. Since the locations of APs follow a homogeneous PPP  ${\Phi}_A$ with density \(\lambda_{\rm A}\), we have 
\begin{align}
|\widetilde{\mathcal{A}}_k|\sim {\rm Poisson} (\lambda_{\rm A} \pi R_1^2 ).	
\end{align}             

To analyze the network performance with a particular choice of fusion weights ${\pmb \zeta}_k$, we choose a uniform random device $k$ in the network and consider its  minimum activity detection probability  $\widetilde{\mathcal{P}}_k({\pmb \zeta}_k)$ as a random variable. We predefine a threshold $P_0$ and define the {\em coverage probability} as 
\begin{align}
{\rm P}_{\rm cov}({\pmb \zeta}_k)&={\rm Pr}(\widetilde{\mathcal{P}}_k({\pmb \zeta}_k)\leq \mathcal{P}_0).
\end{align}
We will use ${\rm P}_{\rm cov}({\pmb \zeta}_k)$ as a performance metric for our network. The coverage probability represents the percentage of devices whose minimum activity detection error probability is smaller than $P_0$. For comparison, we take the following two schemes as benchmarks.

\begin{itemize}
    \item Collocated massive MIMO: All the APs are replaced with a single AP with \(MN=N\lambda_{A}\pi R_0^2\) antennas. This AP is located at the center of a circular area with radius $R_0$. This implies \(\beta_{m,k}=\beta_{k}\sim p_\beta, \theta_{m,(t)}=\theta_{(t)}, \forall m=1,\cdots, M\), and the equal fusing weights are optimal, i.e., ${\pmb \zeta}^{*}_k={\pmb \zeta}^{\#}_k$.
    \item Small-cell system: It is a special case of cell-free IoT, in which the $k$-th device is served only by the nearest AP with index $m^o$, i.e., $\beta_{m^o,k}=\max_{m\in \mathcal{A}_k}\beta_{m,k}$, and  ${\pmb \zeta}^{*}_k={\pmb \zeta}'_k$.
\end{itemize}

For the cell-free stochastic IoT, the coverage probability can be obtained via simulations. For collocated massive MIMO and small-cell systems, we can perform simulations using closed-form expressions derived in the following theorem. 
\begin{theorem}
For collocated massive MIMO system, the coverage probability is
\begin{align}
    {\rm P}_{\rm cov}^{\rm cm}=r_{cm}^2/R_0^2,
\end{align}
where $r_{cm}$ the distance corresponding to the unique solution $\beta_{cm}$ of 
\begin{align}
&\epsilon{{\gamma}\left(NM,NM\frac{\theta_{(t)}}{\beta_{cm}}\log(1+\frac{\beta_{cm}}{\theta_{(t)}})\right)}
+\nonumber\\
&(1-\epsilon){{\Gamma}\left(NM,NM(1+\frac{\theta_{(t)}}{\beta_{cm}})\log(1+\frac{\beta_{cm}}{\theta_{(t)}})\right)}
=\mathcal{P}_0{\Gamma(NM)}.
\end{align}
For small-cell system, the coverage probability is
\begin{align}
    {\rm P}_{\rm cov}^{\rm sc}=1-\exp(-\lambda_A\pi {r}_{sc}^2),
\end{align}
where ${r}_{sc}$ is the distance corresponding to the unique solution $\beta_{sc}$ of 
\begin{align}
&\epsilon{{\gamma}\left(N,N\frac{\theta_{m^{o},(t)}}{\beta_{sc}}\log(1+\frac{\beta_{sc}}{\theta_{m^{o},(t)}})\right)}+\nonumber\\
&(1-\epsilon){{\Gamma}\left(N,N(1+\frac{\theta_{m^{o},(t)}}{\beta_{sc}})\log(1+\frac{\beta_{sc}}{\theta_{m^{o},(t)}})\right)}=\mathcal{P}_0{\Gamma(N)}.
\end{align}
\end{theorem}

\begin{IEEEproof}
For the collocated massive MIMO with \(\beta_{k}\sim p(\beta)\), activity detection error probability \(\widetilde{\mathcal P}_k\) defined in (\ref{P_error_app}) can be simplified as
\begin{align}
{\mathcal P}_k&=\epsilon{{\gamma}\left(NM,NM\frac{\theta_{(t)}}{\beta_k}\log(1+\frac{\beta_k}{\theta_{(t)}})\right)}/{\Gamma(NM)}\nonumber\\
&+(1-\epsilon)
{{\Gamma}\left(NM,NM(1+\frac{\theta_{(t)}}{\beta_k})\log(1+\frac{\beta_k}{\theta_{(t)}})\right)}/{\Gamma(NM)},\nonumber
\end{align}  
which is a monotonic function of $\beta_k$. Thus, the coverage probability is
\begin{align}
    {\rm P}_{\rm cov}^{\rm cm}={\rm Pr}(\beta_k\geq \beta_{cm})=r_{cm}^2/R_0^2,
\end{align}
where the last step follows from the Possion distribution of devices around the collocated AP.

For the small cell system, the activity $\alpha_k$ is determined by the closest AP with index $m^o$. Thus, the activity detection error probability \(\widetilde{\mathcal P}_k\) defined in (\ref{P_error_app}) can be simplified as
\begin{align}
{\mathcal P}_k&=\epsilon{{\gamma}\left(N,N\frac{\theta_{m^{o},(t)}}{\beta_{m^{o},k}}\log(1+\frac{\beta_{m^{o},k}}{\theta_{m^{o},(t)}})\right)}/{\Gamma(N)}\nonumber\\
&+(1-\epsilon){{\Gamma}\left(N,N(1+\frac{\theta_{m^{o},(t)}}{\beta_{m^{o},k}})\log(1+\frac{\beta_{m^{o},k}}{\theta_{m^{o},(t)}})\right)}/{\Gamma(N)},\nonumber
\end{align} 
which is a monotonic function of $\beta_{m^{o},k}$. According to the property of PPP, the CCDF of the distance $r_{m^o,k}$ from the $k$-th device to its nearest AP is
\begin{align}
 {\rm Pr}(r_{m^o,k} \geq x)&=\exp(-\lambda_A\pi x^2).
\end{align}
Thus, the coverage probability for small-cell system is
\begin{align}
    {\rm P}_{\rm cov}^{\rm sc}={\rm Pr}(\beta_{m^o,k}\geq \beta_{sc})=1-\exp(-\lambda_A\pi {r}_{sc}^2).
\end{align}
\end{IEEEproof}

\section{Numerical and Simulation Results}
For a cell-free IoT network, the density of devices is set as  $\lambda_D=637$ { Devices/Km}$^2$. The transmit power of each device, the bandwidth and the power spectral density are assumed to be 23 dBm, 20 MHz, and $-169$ dBm$/$Hz, respectively. Unless stated otherwise, we set $R_0=2$ Km and $\epsilon=0.05$. Similarly as \cite{Xinhua Wang1}, the path-loss coefficient  \(\beta_{l,k}(\mbox{dB})\) is modeled as
\begin{align}
&\beta_{m,k}
=\left\{
\begin{array}{l}
  -\mathcal{L}_0 - 35\log_{10} (r_{m,k}), ~  r_{m,k}>d_1,\\
  -\mathcal{L}_0 - 15\log_{10} (d_1) - 20\log_{10} (d_{0}), ~ r_{m,k} \leq d_0,\\
  -\mathcal{L}_0 - 15\log_{10} (d_1) - 20\log_{10} (r_{m,k}), \mbox{otherwise,}
\end{array}%
\right.\nonumber
 \end{align}
where \(d_0=10\) m, \(d_1=50\) m, and
\begin{align}
\mathcal{L}_0&\triangleq 46.3+33.9\log_{10}(f)-13.82\log_{10}(h_{\text{AP}})\nonumber\\
&-
(1.1\log_{10}(f)-0.7)h_{\text{s}}+(1.56\log_{10}(f)-0.8),\nonumber
\end{align}
where $f=1900$ MHz is the carrier frequency, $h_{\text{AP}}=7$ m and $h_{\text{s}}=1.65$ m are the antenna height of APs and devices, respectively. 
\begin{figure}
\centering \scalebox{1}{\includegraphics[width=\columnwidth]{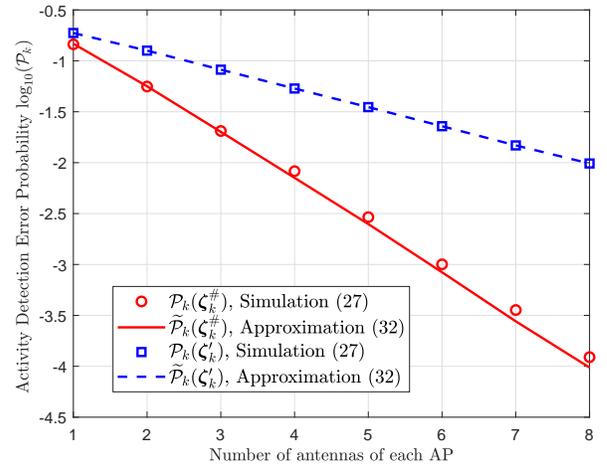}}
\centering \caption{The detection error probability versus the number of antenna at each AP $N$}
\end{figure}

\begin{figure}
\centering \scalebox{1}{\includegraphics[width=\columnwidth]{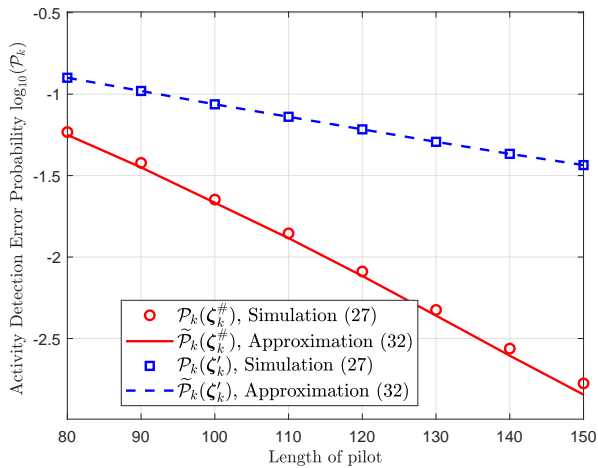}}
\centering \caption{The detection error probability versus the length of pilot}
\end{figure}
\begin{figure}
\centering \scalebox{1}{\includegraphics[width=\columnwidth]{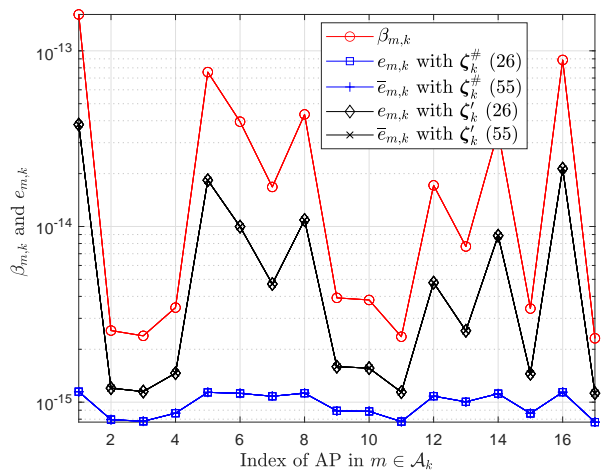}}
\centering \caption{The average mean squared channel estimation error of the $k$-th device}
\end{figure}
We first verify the accuracy of the closed-form expressions for the $k$-th device in Section IV for one realization of path-loss coefficients \(\{\beta_{m,k}\}\). The optimal fusing weights ${\pmb \zeta}^{*}_k$ given in (\ref{opt_fusion}) can be obtained by exhaustive search. To reduce the computation complexity, we use the equal fusing weights ${\pmb \zeta}^{\#}_k$ in (\ref{equ_fusing}) as a sub-optimal substitute of ${\pmb \zeta}^{*}_k$ in the simulations. 

For fixed $\tau=100$, Fig. 2 plots the activity detection error probability of the $k$-th device versus the number of antennas at each AP with different fusing weights ${\pmb \zeta}^{\#}_k$ in (\ref{opt_fusion}) and ${\pmb \zeta}'_k$ in (\ref{sc_fusion}). The benchmark with ${\pmb \zeta}'_k$ can be deemed as the small cell system. The figure shows that the closed-form approximation \(\widetilde{\mathcal P}_{k}({\pmb \zeta}_{k})\) in (\ref{P_error_app}) agrees well with the simulation results given in (\ref{errorpro}), which are obtained through $10,000$ realizations of ${\pmb x}_{m,k}$ and ${\pmb v}_{m,k}$. As predicated by Theorem 2, the activity detection error probability \(\widetilde{\mathcal P}_{k}({\pmb \zeta}^{\#}_{k})\)  decreases rapidly towards zero as $N$ increases. Compared with the benchmark with ${\pmb \zeta}'_k$, the detection error probability is significantly reduced by joint detection with the fusing weight ${\pmb \zeta}^{\#}_k$. 

Fig. 3 shows the detection error probability for the $k$-th device as a function of the length of pilots $\tau$ with $N=3$. It can be seen that both \({\mathcal P}_{k}({\pmb \zeta}^{\#}_{k})\) and \({\mathcal P}_{k}({\pmb \zeta}'_{k})\) decrease significantly as $\tau$ becomes larger. As $\tau=100$ and $N=3$, the activity detection error probability \({\mathcal P}_{k}({\pmb \zeta}^{\#}_{k})\) decreases to about 75$\%$ compared with the benchmark, which indicates the activity detection is relatively more accurate. 

Fig. 4 shows the average mean-squared channel estimation error for the $k$-th device $e_{m,k}, m\in \mathcal{A}_k$ by averaging over $1000$ realizations of random pilots. The closed-form asymptotic expressions \(\overline{e}_k\) in (\ref{channel_app}) agree well with the simulation results \({e}_k\). It is noted that \({e}_{m,k}({\pmb \zeta}^{\#}_{k})\ll {e}_{m,k}({\pmb \zeta}'_{k})\) for any $m$, which implies \({\gamma}_{m,k}({\pmb \zeta}^{\#}_{k})\) 
are close to $\beta_{m,k}$. The philosophy is that the fusing weight ${\pmb \zeta}^{\#}_{k}$ can significantly improve the accuracy of activity detection, and further reduce the mean-squared channel estimation error.

Next, we investigate the impacts of system parameters on the activity detection error probability for $2,000$ realizations of large-scale fading coefficients \(\{\beta_{m,k}\}\). Fig. 5 shows the average activity detection error probability versus the determining radius $R_1$. As $R_1$ increases from 0.5 to 1 km, the average \(\widetilde{{\mathcal P}}_{k}({\pmb \zeta}^{\#}_{k})\) decreases significantly. 
 For $R_1>1.5$ the curve becomes almost flat. 
 It is because the remote APs do not contribute significantly in determining the activity $\alpha_k$ due to the heavy path-loss. Thus, we can select an appropriate determining radius $R_1$ to reduce the overhead of back-haul without losing the accuracy of activity detection.
 
  Fig. 6 plots the CDF of the detection error probability \(\widetilde{{\mathcal P}}_{k}({\pmb \zeta}^{\#}_{k})\) for 2000 realizations of large-scale fading coefficients \(\{\beta_{m,k}\}\). For cell-free IoT networks, it is interesting to investigate the impact of density of APs. To do this, we fix $N\lambda_A=10$, which implies the average total number of antennas in an unit area is fixed, and use $\lambda_A=2$ and 5. By adopting the fusing weight ${\pmb \zeta}^{\#}_k$, the detection error probabilities can be significantly reduced compared with fusing weights ${\pmb \zeta}'_k$. We also note that the 95$\%$ likely performance of  \(\widetilde{{\mathcal P}}_{k}({\pmb \zeta}^{\#}_{k})\) with $\lambda_A=5$ is 2 $\%$, while the 95$\%$ likely performance of  \(\widetilde{{\mathcal P}}_{k}({\pmb \zeta}^{\#}_{k})\) with $\lambda_A=2$ is 10 $\%$. Therefore, larger density of APs allows to improve the accuracy of activity detection when giving the total number of antennas. This observation is expected since the larger number of APs, the higher chances that some of them are located close to an active user. 

\begin{figure}
\centering \scalebox{1}{\includegraphics[width=\columnwidth]{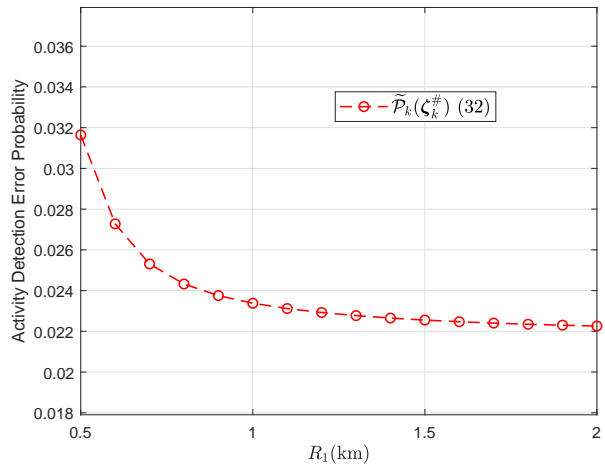}}
\centering \caption{The average detection error probability versus the determining radius $R_1$}
\end{figure}
\begin{figure}
\centering \scalebox{1}{\includegraphics[width=\columnwidth]{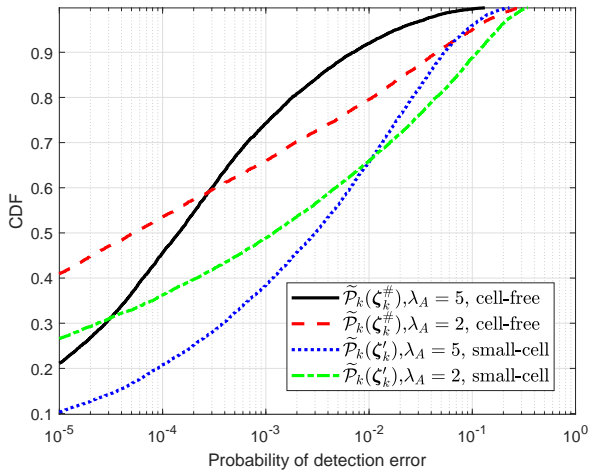}}
\centering \caption{The CDF of detection error probability with fixed $N\lambda_A=10$}
\end{figure}

Finally, we compare the coverage probability of three systems, namely cell-free IoT, collocated massive MIMO, and small-cell IoT for $2,000$ realizations of path-loss coefficients \(\{\beta_{l,k}\}\). Fig. 7 shows the coverage probabilities with $P_{0}=0.02$ versus the number of antennas $N$ at each AP. With the increase of $N$, all the coverage probability increase significantly. The figure reveals that the closed-form expressions for the small cell system and the collocated massive MIMO match well with the simulation results by averaging over $2,000$ realizations of path-loss coefficients. The coverage probability of cell-free networks with ${\pmb \zeta}_k^{\#}$ is much higher than the coverage probabilities of small-cell systems and co-located MIMO.

\begin{figure}
\centering \scalebox{1}{\includegraphics[width=\columnwidth]{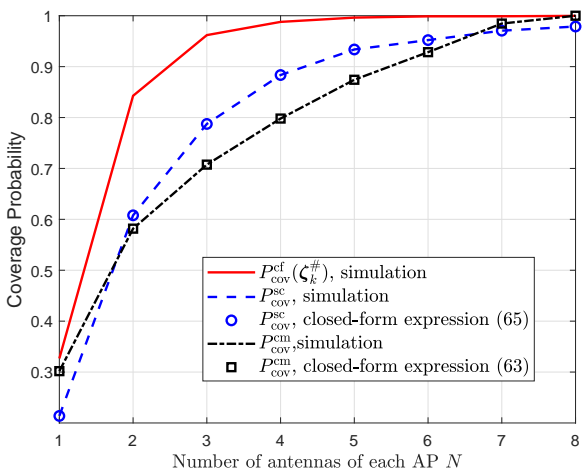}}
\centering \caption{The coverage probability versus the number of antenna at each AP $N$}
\end{figure}
\section{Conclusion}

We investigated the activity detection and channel estimation for a cell-free IoT network with massive random access. Focusing on the distributed and user-centric architecture, we proposed a two-stage approach based on the vector AMP algorithm. We derived the 
activity detection error probability and the mean-squared channel estimation error in closed-form as metrics for a typical device. We defined the coverage probability as a performance metric for the stochastic network. Numerical results show that the coverage probability  of the cell-free IoT network can be significantly improved compared with the collocated massive MIMO and small-cell counterparts.

\section*{Appendices}
\subsection{Useful Lemmas}
\begin{lemma}
Let ${\pmb A}\in \mathbb{C}^{\tau\times \tau}$ be a Hermitian invertible matrix. Then, for any vector ${\pmb x}\in \mathbb{C}^{\tau}$ and any scalar $a\in \mathbb{C} $ such that ${\pmb A}+a{\pmb x}{\pmb x}^H$ is invertible, we have
\begin{align}
   {\pmb x}^H\left({\pmb A}+a{\pmb x}{\pmb x}^H\right)^{-1}=\frac{{\pmb x}^H{\pmb A}^{-1}}{1+a{\pmb x}^H{\pmb A}^{-1}{\pmb x}}, \label{matrix_inverse_1}\\
     \left({\pmb A}+a{\pmb x}{\pmb x}^H\right)^{-1}{\pmb x}=\frac{{\pmb A}^{-1}{\pmb x}}{1+a{\pmb x}^H{\pmb A}^{-1}{\pmb x}}. \label{matrix_inverse_2}
\end{align}
\end{lemma}

\begin{lemma}
Let ${\pmb A}\in \mathbb{C}^{\tau\times \tau}$, and  ${\pmb x},{\pmb y}\sim \mathcal{CN}({\pmb 0}, \frac{1}{\tau}{\pmb I}_{\tau})$. Assume that ${\pmb A}$ has uniformly bounded spectral norm (with respect to $\tau$) and that ${\pmb x}$ and ${\pmb y}$ are mutually independent and independent of ${\pmb A}$, we have
\begin{align}
   & {\pmb x}^H{\pmb A}{\pmb x}-{{\rm tr}{\pmb A}}/{\tau}\mathop{-\!\!\!\!-\!\!\!\!-\!\!\!\!-\!\!\!\!-\!\!\!\!-\!\!\!\!\longrightarrow}\limits_{\tau\rightarrow\infty}^{{\rm a}.{\rm s}.}0, \label{xAx}\\
    &{\pmb x}^H{\pmb A}{\pmb y}\mathop{-\!\!\!\!-\!\!\!\!-\!\!\!\!-\!\!\!\!-\!\!\!\!-\!\!\!\!\longrightarrow}\limits_{\tau\rightarrow\infty}^{{\rm a}.{\rm s}.}0.\label{xAy}
\end{align}
\end{lemma}
\begin{lemma}
Let ${\pmb x},~{\pmb y}\in \mathbb{C}^{N}$ be two independent random variables satisfying $p_{\pmb X}({\pmb x})=p_{\pmb X}(-{\pmb x})$ and  $p_{\pmb Y}({\pmb y})=p_{\pmb Y}(-{\pmb y})$, and $f({\pmb x},{\pmb y})$ be an even function. Then we have
\begin{align}
    \mathbb{E}_{{\pmb x},~{\pmb y}}[f({\pmb x},~{\pmb y}){\pmb x}^H{\pmb y}]=0.
\end{align}
\end{lemma}
\begin{lemma}
Let $x,~y\in \mathbb{C}$ are two i.i.d. random variables, and $f(x,y)$ is a symmetric function satisfying $$f(x=x_0,y=y_0)=f(x=y_0,y=x_0),$$ for any $x_0,~y_0\in \mathbb{C}$, we have
\begin{align}
    \mathbb{E}_{x,y}[f(x,y)x^Hx]=\mathbb{E}_{x,y}[f(x,y)y^Hy].
\end{align}
\end{lemma}

\begin{lemma}[{Properties of Gamma Functions}\cite{Abramowitz}]
\begin{enumerate}
    \item As $\Re(s)>0$, we have
    \begin{align} {\gamma}(s,x)+{\Gamma}(s,x)={\Gamma}(s).\label{Gamma_pro1}
\end{align}
  \item As ${\displaystyle s\to \infty }$ and $n\in \mathbb{C}$, we have the following asymptotic approximation
    \begin{align} \mathop{\lim}_{s\rightarrow \infty} \frac{\Gamma(s+n)}{\Gamma(s)s^n}=1. \label{Gamma_pro2}
\end{align}
  \item As ${\displaystyle s\to \infty }$, we have the following asymptotic approximation
    \begin{align} \mathop{\lim}_{s\rightarrow \infty}\Gamma(s+1)=\sqrt{2\pi s}s^se^{-s}.\label{Gamma_pro3} 
\end{align}
  \item  As $\Re(s)>0$, the power series expansion of the lower incomplete gamma function is
    \begin{align} {\gamma}(s,x)= x^{s}e^{-x}\Gamma(s)\sum\nolimits_{i=0}^{\infty}\frac{x^{i}}{\Gamma(s+i+1)}. \label{Gamma_pro4}
\end{align}
 \item As ${\sqrt{s}}/{(x-s)}\rightarrow 0$, the asymptotic expansion
of the upper incomplete gamma function is given by \cite{Tricomi, W. Gautschi, M. Sharif}
\begin{align}\label{incom_gamma_up}
\Gamma(s, x)&=\frac{e^{-x}x^{s}}{x-s+1}\left\{1-\frac{s-1}{(x-s+1)^{2}}\right.\nonumber\\
&\left.+\frac{2(s-1)}{(x-s+1)^{3}}+O\left(\frac{(s-1)^{2}}{(x-s+1)^{4}}\right)\right\}.
\end{align}
\end{enumerate}
\end{lemma}
\subsection{Proof of Theorem 1}
Since \({\pmb x}_{m,k}\sim p_{{\pmb x}_{m,k}}=(1-\epsilon)\delta_{\bf 0}+\epsilon p_{{\pmb g}_{m,k}}\), where $p_{{\pmb g}_{m,k}}$ is the PDF of \({\pmb g}_{m,k}\sim \mathcal{CN}({\pmb 0},{\beta}_{m,k}{\pmb I}_N)\), the initial state ${\pmb \Theta}_{m,(0)}$ in (\ref{state evolution_initial}) can be simplified into 
\begin{align}\label{theta_0}
    {\pmb \Theta}_{m,(0)}=\theta_{(m,0)}{\pmb I}_N,
 \end{align}
 where \(\theta_{m,(0)}\) is given in ({\ref{state_eve_simple}\text{a}}).

\begin{figure*}
  \begin{align}\label{SE}
\pmb{\Theta}_{m,(t+1)}&=\frac{\pmb{I}}{E}+
	\frac{|\mathcal{D}_m|}{\tau}\mathbb{E}_{\mathcal{D}_m}\left\{[\eta(\widehat{\pmb{X}}_{m,k,(t)})-{\pmb{X}}_{m,k}][\eta(\widehat{\pmb{X}}_{m,k,(t)})-{\pmb{X}}_{m,k}]^H\right\}\nonumber\\
	&\overset{(a)}{=}\frac{\pmb{I}}{E }+
	\frac{|\mathcal{D}_m|}{\tau} \mathbb{E}_{\mathcal{D}_m}\left\{\left[\xi_{m,k,(t)}\widehat{{\pmb X }}_{m,k,(t)}-{\pmb{X}}_{m,k}\right] \left[\xi_{m,k,(t)}\widehat{{\pmb X }}_{m,k,(t)}-{\pmb{X}}_{m,k}\right]^H\right\}\nonumber\\
	&\overset{(b)}{=}\frac{\pmb{I}}{E }+
	\frac{|\mathcal{D}_m|}{\tau} \mathbb{E}_{\mathcal{D}_m}\left\{\left(\xi_{m,k,(t)}-1\right)^2{{\pmb X }}_{m,k}{{\pmb X }}_{m,k}^H+\left(\xi_{m,k,(t)}\theta_{(t)}\right)^2{{\pmb V }_{m,k}}{{\pmb V }}_{m,k}^H\right.\nonumber\\
	&\left.\qquad\qquad\quad+\left(\xi_{m,k,(t)}-1\right)\xi_{m,k,(t)}\theta_{(t)}\left({{\pmb X }}_{m,k}{{\pmb V }}_{m,k}^H+{{\pmb V }}_{m,k}{{\pmb X }}_{m,k}^H\right)\right\}\nonumber\\
		&\overset{(c)}{=}\frac{\pmb{I}}{E }+
	\frac{|\mathcal{D}_m|}{\tau} \mathbb{E}_{\mathcal{D}_m}\left\{\left(\xi_{m,k,(t)}-1\right)^2{{\pmb X }}_{m,k}{{\pmb X }}_{m,k}^H+\left(\xi_{m,k,(t)}\theta_{(t)}\right)^2{{\pmb V }_{m,k}}{{\pmb V }}_{m,k}^H\right\},
	 \end{align}
	 \hrulefill
\end{figure*}
For given \({\pmb \Theta}_{(t)}=\theta_{m,(t)}{\pmb I}_N\), using (\ref{eqn:state evolution}), we have (\ref{SE}) shown at the top of this page.
Step (a) is obtained according the definition of MMSE denoiser given in (\ref{MMSE_denoiser_def}) and assumption \({\pmb \Theta}_{m,(t)}=\theta_{m,(t)}{\pmb I}_N\), step (b) is obtained according to the signal model (\ref{noise_model}), step (c) is obtained using Lemma 5 since $\xi_{m,k,(t)}$ is an even function w.r.t. ${\pmb X }_{m,k}$ and ${\pmb V }_{m,k}$ according to  (13c), (12a)-(12c), and (9).

Define 
\begin{align}
{\pmb B}&=\mathbb{E}_{\mathcal{D}_m}\left[\left(\xi_{m,k,(t)}-1\right)^2{{\pmb X }}_{m,k}{{\pmb X }}_{m,k}^H \right.\nonumber\\
&\qquad \qquad \qquad \left.+\xi_{m,k,(t)}^2\theta_{m,(t)}^2{{\pmb V }_{m,k}}{{\pmb V }}_{m,k}^H\right].
\end{align}
For the non-diagonal entity ${\pmb B}_{(i,j)}$ with $i\neq j$, we have 
\begin{align}\label{nondiag_ele}
{\pmb B}_{(i,j)}&=\mathbb{E}_{\mathcal{D}_m}\left[\left(\xi_{m,k,(t)}-1\right)^2\left[{{\pmb X }}_{m,k}\right]_{(i)}\left[{{\pmb X }}_{m,k}^*\right]_{(j)}\right.\nonumber\\
&\left.+\xi_{m,k,(t)}^2\theta_{m,(t)}^2\left[{{\pmb V }_{m,k}}\right]_{(i)}\left[{{\pmb V }}_{m,k}^*\right]_{(j)}\right]=0\end{align}
where the last step follows from Lemma 5 since $\left[{\pmb X }_{m,k}\right]_{(i)}$ and $\left[{\pmb V }_{m,k}\right]_{(i)}$ are independent of $\left[{\pmb X }_{m,k}\right]_{(j)}$ and $\left[{\pmb V }_{m,k}\right]_{(j)}$, respectively. 
For the diagonal entities, according to Lemma 6, we have 
$$\left[{\pmb B}\right]_{(i,i)}=\left[{\pmb B}\right]_{(j,j)},~\forall i,j=1,\cdots,N,$$
since $\xi_{m,k,(t)}$ is a symmetric function of the entities of ${\pmb X }_{m,k}$ and ${\pmb V }_{m,k}$. Hence, we have
\begin{align}\label{diag_ele}
    \left[{\pmb B}\right]_{(i,i)}=\frac{{\rm tr}[{\pmb B}]}{N}, ~\forall i=1,\cdots,N.
\end{align}
According to (\ref{nondiag_ele}) and (\ref{diag_ele}), $\pmb{\Theta}_{m,(t+1)}$ can be further simplified as
\begin{align}\label{theta_t}
    {\pmb \Theta}_{m,(t+1)}=\theta_{m,(t+1)}{\pmb I}_N,
 \end{align}
with $\theta_{m,(t+1)}$ defined in (\ref{state_eve_simple}\text{b}). Combining (\ref{theta_0}) and (\ref{theta_t}), we complete the proof.
\subsection{Proof of Lemma 1}
For $c\in (0,1)$, we have
\begin{align}
\mathop{\lim}_{s\rightarrow +\infty}\frac{{\gamma}(s,c s)}{\Gamma(s)}\overset{(\ref{Gamma_pro4})}{=}& \mathop{\lim}_{s\rightarrow +\infty}\frac{(c s)^{s}}{e^{(c s)}}\sum_{i=0}^{\infty}\frac{(c s)^{i}}{\Gamma(s+i+1)}\nonumber\\
\overset{(\ref{Gamma_pro2})}{\geq} &\mathop{\lim}_{s\rightarrow +\infty}\frac{(c s)^{s}}{e^{(c s)}}\sum_{i=0}^{\infty}\frac{(c s)^{i}}{\Gamma(s)(s)^{i+1}}\nonumber\\
=&\mathop{\lim}_{s\rightarrow +\infty}\frac{(c s)^{s}}{s\Gamma(s)e^{(c s)}}\sum_{i=0}^{\infty}(c)^{i}\nonumber\\
=&\mathop{\lim}_{s\rightarrow +\infty}\frac{(c s)^{s}e^{(-c s)}}{s(1-c)\Gamma(s)}\nonumber\\
\overset{(\ref{Gamma_pro2})}{=}&\mathop{\lim}_{s\rightarrow +\infty}\frac{(c s)^{s}e^{(-c s)}}{(1-c)\Gamma(s+1)}\nonumber\\
\overset{(\ref{Gamma_pro3})}{=} &\mathop{\lim}_{s\rightarrow +\infty} \frac{(c s)^{s}e^{s}}{\sqrt{2\pi s}s^{s}(1-c)e^{(c s)}}\nonumber\\
= &\mathop{\lim}_{s\rightarrow +\infty}\frac{1}{(1-c)\sqrt{2\pi s}}\left(\frac{c e}{e^{c}}\right)^{s}\nonumber\\
=&0,
\end{align}
where the last step is based on the fact $$0<\frac{c e}{e^{c}}<1,~\mbox{as}~0\leq c\leq 1.$$

Consider now the case $c'>1$. By substituting $x=c's$ into (\ref{incom_gamma_up}) and taking the limit, we obtain 
\begin{align}
\mathop{\lim}_{s\rightarrow +\infty} \frac{{\Gamma}(s,c's)}{\Gamma(s)}= &\mathop{\lim}_{s\rightarrow +\infty}\frac{(c's)^{s}e^{(-c's)}}{(c'-1)s\Gamma(s)}\cr
\overset{(\ref{Gamma_pro2})}{=} &\mathop{\lim}_{s\rightarrow +\infty}\frac{(c's)^{s}e^{(-c's)}}{(c'-1)\Gamma(s+1)}\cr
\overset{(\ref{Gamma_pro3})}{=} &\mathop{\lim}_{s\rightarrow +\infty}\frac{(c's)^{s}e^{(-c's)}e^{s}}{(c'-1)\sqrt{2\pi s}s^{s}}\cr
= &\mathop{\lim}_{s\rightarrow +\infty}\frac{1}{(c'-1)\sqrt{2\pi s}}\left(\frac{c' e}{e^{c'}}\right)^{s}\cr
=&0,
\end{align}
where the last equality is based on the fact $$0<\frac{c e}{e^{c}}<1,~\mbox{as}~ c> 1.$$

\end{document}